\documentclass[twocolumn,nofootinbib,floatfix,
superscriptaddress]{revtex4}        
\usepackage{graphicx}
\usepackage{epsfig}
\usepackage{bm}
\usepackage[T1]{fontenc}
\usepackage[latin9]{inputenc}
\usepackage{amssymb}
\usepackage{float}
\usepackage{amsmath}
\usepackage{dcolumn}
\usepackage{cancel}
\usepackage[colorlinks]{hyperref}
\usepackage[usenames,dvipsnames]{color}
\hypersetup{
    breaklinks=true,
    pdfstartview={FitH},    
    colorlinks=true,       
    linkcolor=blue,          
    citecolor=red,        
    filecolor=magenta,      
    urlcolor=blue,           
    anchorcolor=green,      
    linktocpage=true
}

\newcommand{\be}{\begin{equation}}
\newcommand{\ee}{\end{equation}}

\begin{document}
\title{Note on agegraphic dark energy inspired by modified Barrow entropy}

\author{A. Sheykhi}
\email{asheykhi@shirazu.ac.ir} \affiliation{Department of Physics,
College of Sciences, Shiraz University, Shiraz 71454, Iran}
\affiliation{Biruni Observatory, College of Sciences, Shiraz
University, Shiraz 71454, Iran}

\author{S. Ghaffari}
\email{sh.ghaffari@maragheh.ac.ir}
\affiliation{Research Institute for Astronomy and Astrophysics of Maragha (RIAAM), University of Maragheh, P.O. Box 55136-553, Maragheh, Iran}

\begin{abstract}
We revisit agegraphic dark energy (ADE) model when the entropy
associated with the apparent horizon is in the form of Barrow
entropy, $S\sim A^{1+\delta/2}$, where $0\leq\delta\leq1$
indicates the amount of the quantum-gravitational deformation
effects of the horizon. The modification to the entropy
expression, not only change the energy density of ADE, but also
modifies the Friedmann equations due to thermodynamics-gravity
conjecture. Based on this, we investigate the cosmological
consequences of ADE through modified Barrow cosmology and disclose
the effects of Barrow exponent $\delta$ on the evolutions of the
cosmological parameters. We observe that, depending on the values
of $\delta$, the transition from early decelerated phase to the
late time accelerated phase occurs, and the equation of state
(EoS) parameter $ w_{de} $ varies from quintessence $
-1<w_{de}<-1/3 $ to the phantom regime $ (w_{de}<-1)$. When
$\delta=0$, all results of ADE in standard cosmology are restored.


\end{abstract}
\maketitle
\section{Introduction\label{Int}}
According to the thermodynamics-gravity conjecture, one can
rewrite the Friedmann equations of Friedmann-Robertson-Walker
(FRW) universe in the form of the first law of thermodynamics on
the apparent horizon and vice versa \cite{Cai1,Cai2,
Cai3,Akbar1,Akbar2,Akbar3,
Sheykhi1,Sheykhi3,Sheykhi4,Sheykhi5,SheyECFE}. For several
reasons, this correspondence is quite interesting. First, it
supports the holography idea, since the Friedmann equations
describe the evolution of the universe in the bulk, while the
first law of thermodynamics is written on the boundary. Second, it
provides a way to derive the field equations of gravity using
thermodynamic arguments. Third, in some cases, where the entropy
expression is not well established, it can be useful to derive the
entropy expression associated with the horizon
\cite{Sheykhi3,Sheykhi4}.

On the other side, dark energy (DE) puzzle, as responsible energy
for describing the acceleration of the cosmic expansion, has been
one of the attractive field of research, for both theoretical and
experimental cosmologists. One of the well-known candidates of DE
is the so called holographic DE  which is based on the holographic
principle developed by Fischler and Susskind
\cite{Suss1,Suss2,Coh}. Another interesting model is the so called
ADE model which suggested by Cai in 2007 \cite{CaiADE}. This model
is based on the uncertainty relation of quantum mechanics together
with the gravitational effect in general relativity. Accordingly,
the energy density of metric fluctuations of the Minkowski
spacetime is given by  $\rho_{de} \sim {m^2_p}/{t^2},$ where
$m_{p}$ is the reduced Planck mass \cite{Maz}. In this model the
observed DE originates from the spacetime and matter field
fluctuations in the universe \cite{Wei1,Wei2}. The ADE models have
received a lot of attentions in the literatures (see e.g.
\cite{Cui,Kim,sheyADE1,sheyADE2,Setare} and checked with
observations \cite{Wei3}). Other studies on ADE models can be
carried out in \cite{JKS,Majid,Pan,Kumar}.

According to the new proposal of J.D. Barrow, the horizon of a
black hole may has a fractal structure and the corresponding area
of the horizon cloud increase due to the quantum-gravitational
deformation \cite{Barrow}. In this scenario, the area law of
entropy get modified as $S_{h} \sim A^{ 1+\delta/2} $ where
$\delta$ quantifies the quantum-gravitational deformation. In the
cosmological setup, the effects of Barrow entropy on the cosmic
evolution have been explored from different viewpoints. In this
regards, a new  holographic DE model based on Barrow entropy has
been proposed \cite{Emm1,Ana}. A cosmological scenarios based on
Barrow entropy was investigated in \cite{Emm2}, where it was shown
that new extra terms that constitute an effective energy density
sector are appeared in the Friedmann equations. In another
approach, the modified Friedmann equations through Barrow entropy
and its cosmological consequences were explored in
\cite{SheB1,SheB2,SheB3}, where it was shown that the geometry
part of the Friedmann equations get modified due to the
corrections to the entropy expression. It is worth noting that the
exponent $\delta$ in Barrow entropy, cannot reproduce any term
which may play the role of DE and one still needs to take into
account the DE (cosmological constant) component in the Friedmann
equations to reproduce the accelerated universe
\cite{SheB2,SheB3}. On the other hand, it was recently proven that
Barrow entropy as well as any other known entropy (Tsalis, Renyi,
Kaniadakis, etc) is just sub-case of generalized entropy
expression introduced in \cite{Odin1,Odin2}. Other studies on the
cosmological consequences of the Barrow entropy can be carried out
in \cite{Emm3,Abr1,Mam,Abr2,Bar2,Sri,Das,Pra,Odin3,Odin4}.

The modified ADE model when the entropy is in the form of Barrow
entropy, $S_{h} \sim A^{ 1+\delta/2} $, has been explored in
\cite{BADE}. However, the authors of \cite{BADE} only modify the
energy density of ADE, while they still use the standard Friedmann
equations. This is indeed inconsistent with the
thermodynamics-gravity conjecture. Indeed, any modification to the
entropy expression, not only change the energy density of ADE, but
also modifies the Friedmann equations describing the background
evolution. Our work differs from  \cite{BADE} in that we consider
modification to both energy density as well as Friedmann equations
due to the Barrow correction to the area law of entropy. In the
light of all mentioned above, it becomes obvious that the
investigation on the ADE models in the context of modified Barrow
cosmology is well motivated. In particular, we would like to
disclose the effects of Barrow exponent $\delta$ on the
cosmological parameters in the late time when the ADE is
dominated.

The plan of the work is as follows. In section  \ref{Fri}, we
review constructing the modified Friedmann equations through
Barrow entropy using the thermodynamics-gravity conjecture. In
section \ref{ORI}, we study the cosmological consequences of the
ADE in the context of modified Barrow cosmology. In section
\ref{NEW}, we consider the new model of ADE while the time scale
is chosen to be the conformal time instead of the age of the
universe. The last section is devoted to conclusions and
discussions.
\section{Modified Friedmann equations through Barrow entropy\label{Fri}}
Consider a homogeneous and isotropic FRW universe which is
described by the line elements
\begin{equation}
ds^2={h}_{\mu \nu}dx^{\mu}
dx^{\nu}+\tilde{r}^2(d\theta^2+\sin^2\theta d\phi^2),
\end{equation}
where $\tilde{r}=a(t)r$, $a(t)$ is the scale factor, $x^0=t,
x^1=r$, and $h_{\mu \nu}$=diag $(-1, a^2/(1-kr^2))$ represents the
two dimensional metric. Here $k=0,\pm 1$ represent the curvature
of the three dimensional space. We take apparent horizon as the
boundary of spacetime, with radius \cite{Akbar2}
\begin{equation}
\label{radius} \tilde{r}_A=\frac{1}{\sqrt{H^2+k/a^2}},
\end{equation}
where $H=\dot{a}/a$ stands for the Hubble parameter which measures
the rate of the universe expansion. The temperature of the
apparent horizon is given by \cite{Akbar2}
\begin{equation}\label{T}
T_h=-\frac{1}{2 \pi \tilde r_A}\left(1-\frac{\dot {\tilde
r}_A}{2H\tilde r_A}\right).
\end{equation}
We assume, the energy momentum tensor of the universe is in the
form of perfect fluid,$
T_{\mu\nu}=(\rho+p)u_{\mu}u_{\nu}+pg_{\mu\nu},$ where $\rho$ and
$p$ are, respectively, the energy density and pressure of the
matter inside the Universe. The total energy content of the
universe satisfies the conservation equation,
$\nabla_{\mu}T^{\mu\nu}=0$, which yields the continuity equation,
\begin{equation}\label{Cont}
\dot{\rho}+3H(\rho+p)=0.
\end{equation}
Since our universe is expanding thus we have a work term in the
first law of thermodynamics. The work density is given by
\cite{Hay2}
\begin{equation}\label{Work}
W=-\frac{1}{2} T^{\mu\nu}h_{\mu\nu},\ \ \Rightarrow \ \
W=\frac{1}{2}(\rho-p).
\end{equation}
We propose the first law of thermodynamics holds on the apparent
horizon
\begin{equation}\label{FL}
dE = T_h dS_h + WdV,
\end{equation}
where $V=\frac{4\pi}{3}\tilde{r}_{A}^{3}$ is the volume enveloped
by a 3-dimensional sphere, and $T_{h}$ and $W$ are given by Eqs.
(\ref{T}) and (\ref{Work}). In the above expression $S_{h}$ is the
entropy associated with the apparent horizon. Taking differential
of total energy inside the apparent horizon, $E=\rho V$, yields
\begin{equation}
\label{dE} dE=4\pi\tilde {r}_{A}^{2}\rho d\tilde {r}_{A}-4\pi H
\tilde{r}_{A}^{3}(\rho+p) dt.
\end{equation}
where we have also used the continuity equation (\ref{Cont}). The
Barrow entropy associated to the apparent horizon is given by
\cite{Barrow}
\begin{eqnarray}\label{SB}
S_{h}= \left(\frac{A}{A_{0}}\right)^{1+\delta/2},
\end{eqnarray}
where $A=4\pi \tilde{r}_{A}^{2}$ is apparent horizon area and
$A_0$ is the Planck area. The exponent $\delta$ ranges as
$0\leq\delta\leq1$ and stands for the amount of the
quantum-gravitational deformation effects \cite{Barrow}. When
$\delta=0$, the area law is restored and $A_{0}\rightarrow 4G$,
while $\delta=1$ represents the most intricate and fractal
structure of the horizon. Using the thermodynamics-gravity
conjecture, the differential form of the Friedmann equation
derived from the first law of thermodynamics (\ref{FL}), based on
Barrow entropy, is given by (see \cite{SheB1} for details)
\begin{equation} \label{FriedB}
-\frac{2+\delta}{2\pi A_0
}\left(\frac{4\pi}{A_0}\right)^{\delta/2} \frac{d\tilde
{r}_{A}}{\tilde {r}_{A}^{3-\delta}}=
 \frac{d\rho}{3}.
\end{equation}
After integration, we find the modified first Friedmann equation
in Barrow cosmology as
\begin{equation} \label{Fried0B}
\left(H^2+\frac{k}{a^2}\right)^{1-\delta/2} = \frac{8\pi G_{\rm
eff}}{3} \rho+\frac{\Lambda}{3},
\end{equation}
where $\Lambda$ is a constant of integration which can be
interpreted as the cosmological constant, and  $G_{\rm eff}$
stands for the effective Newtonian gravitational constant,
\begin{equation}\label{Geff}
G_{\rm eff}\equiv \frac{A_0}{4} \left(
\frac{2-\delta}{2+\delta}\right)\left(\frac{A_0}{4\pi
}\right)^{\delta/2}.
\end{equation}
If we define $\rho_{\Lambda}={\Lambda}/(8\pi G_{\rm eff})$, Eq.
(\ref{Fried0B}), can be rewritten as
\begin{equation} \label{Fried1B}
\left(H^2+\frac{k}{a^2}\right)^{1-\delta/2} = \frac{8\pi G_{\rm
eff}}{3}(\rho+\rho_{\Lambda}).
\end{equation}
Combining the modified Friedmann equation (\ref{Fried1B}) with the
continuity equation (\ref{Cont}), we arrive at
\begin{eqnarray}
&&(2-\delta)\frac{\ddot{a}}{a}
\left(H^2+\frac{k}{a^2}\right)^{-\delta/2}+(1+\delta)\left(H^2+\frac{k}{a^2}\right)^{1-\delta/2}
\nonumber
\\
&&=-8\pi G_{\rm eff}(p+p_{\Lambda}),\label{Fried2B}
\end{eqnarray}
where $p_{\Lambda}=-{\Lambda}/(8\pi G_{\rm eff})$. This is the
second modified Friedmann equation governing the evolution of the
universe based on Barrow entropy. In the limiting case where
$\delta=0$ ($G_{\rm eff}\rightarrow G$), Eqs. (\ref{Fried1B}) and
(\ref{Fried2B}) reduce to the Friedmann equation in standard
cosmology.

As usual, we define the density parameters as
\begin{eqnarray}
\Omega_{i}=\frac{\rho_{i}}{\rho_{c}}, \  \   \
\rho_{c}=\frac{3H^{2-\delta}}{8\pi G_{\rm eff}}. \  \
\end{eqnarray}
Therefore, in terms of the density parameters, the first Friedmann
equation (\ref{Fried1B}) can be written as
\begin{eqnarray}
\Omega_m+ \Omega_{\Lambda}=(1+\Omega_k)^{1-\delta/2}.
\end{eqnarray}
where the curvature density parameter is defined as usual,
$\Omega_k=k/(a^2H^2)$.

Now we are going to show that the modified Friedmann equations
derived in Eq. (\ref{Fried1B}) can describe the late time
accelerated expansion only in the presence of DE (cosmological
constant). For a flat universe filled with pressureless matter
($p=p_m=0$) and cosmological constant, we have $\Omega_m+
\Omega_{\Lambda}=1.$ The total EoS parameter can be written
\begin{eqnarray}
&&\omega_{\rm
t}=\frac{p_{\Lambda}}{\rho_m+\rho_{\Lambda}}=-\frac{1}{\rho_m/\rho_{\Lambda}+1}, \\
&&\Rightarrow\nonumber \omega_{\rm t}
(z)=-\frac{\Omega_{\Lambda,0}}{(1-\Omega_{\Lambda ,0})
(1+z)^3+\Omega_{\Lambda,0}},
\end{eqnarray}
where $p_{\Lambda}/\rho_{\Lambda}=-1$,
$\rho_{\Lambda}=\rho_{\Lambda,0}$, and $\rho_m=\rho_{m,0}
(1+z)^{3}$. If we take $\Omega_{\Lambda,0} \simeq 0.7 $ and
$\Omega_{m,0} \simeq 0.3$, we have
\begin{eqnarray}
&&\omega_{\rm t} (z)=-\frac{0.7}{0.7+0.3 (1+z)^3}.
\end{eqnarray}
At the present time where $z\rightarrow 0$, we have $\omega_{\rm
t}=-0.7$, while at the early universe where $z\rightarrow \infty$,
we get $\omega_{\rm t}=0$. This implies that at the early stages,
the universe undergoes a decelerated phase while at the late time
it experiences an accelerated phase.
\begin{figure}[htp]
    \begin{center}
        \includegraphics[width=8.7cm]{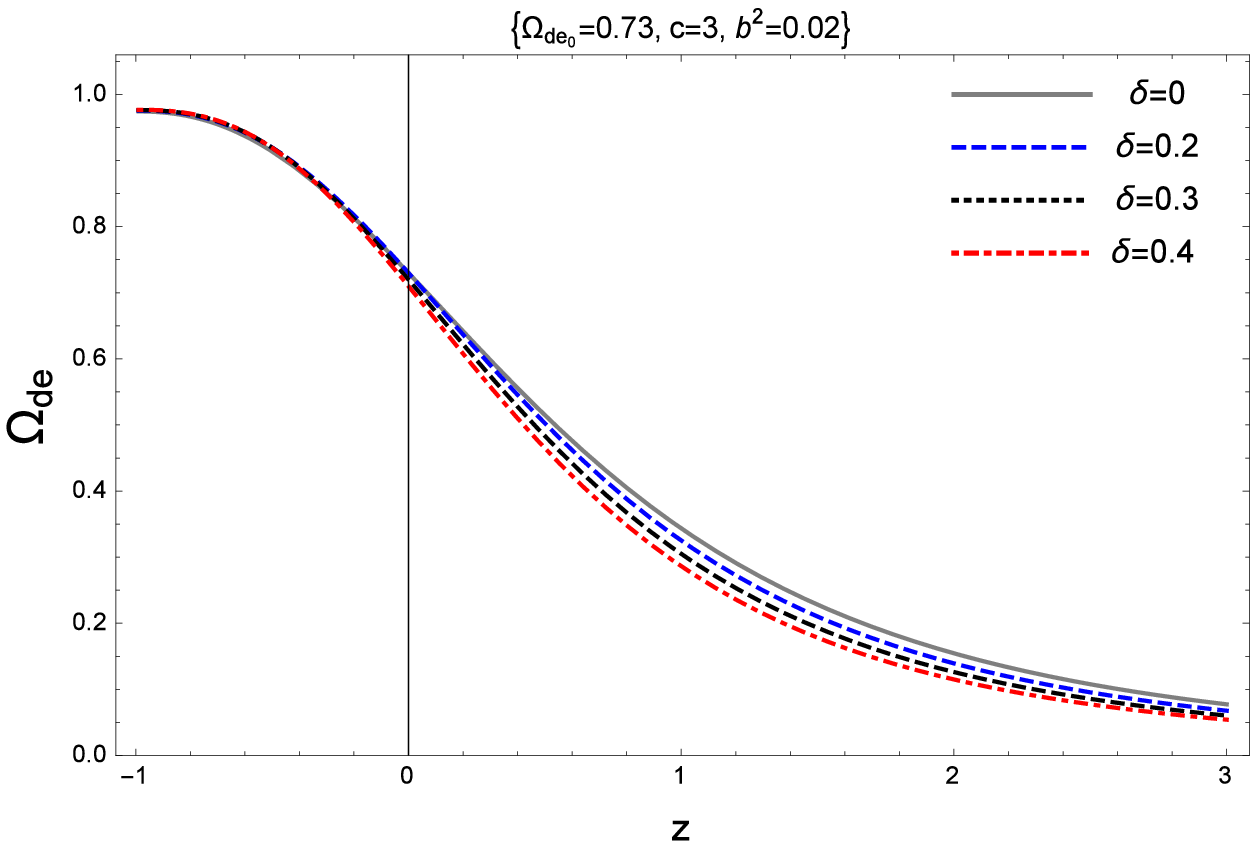}\text{(a)}
        \vspace{1mm}
        \includegraphics[width=8.7cm]{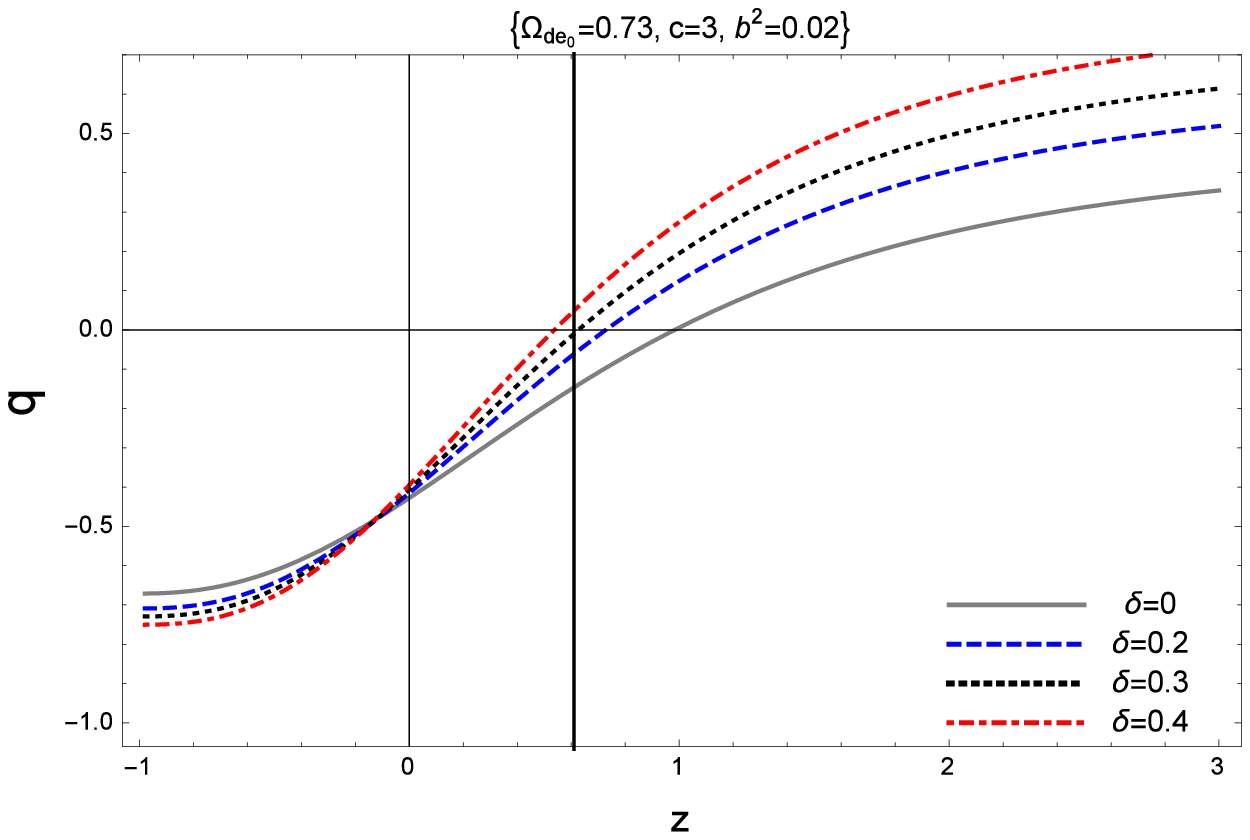}\text{(b)}
        \vspace{1mm}
        \includegraphics[width=8.7cm]{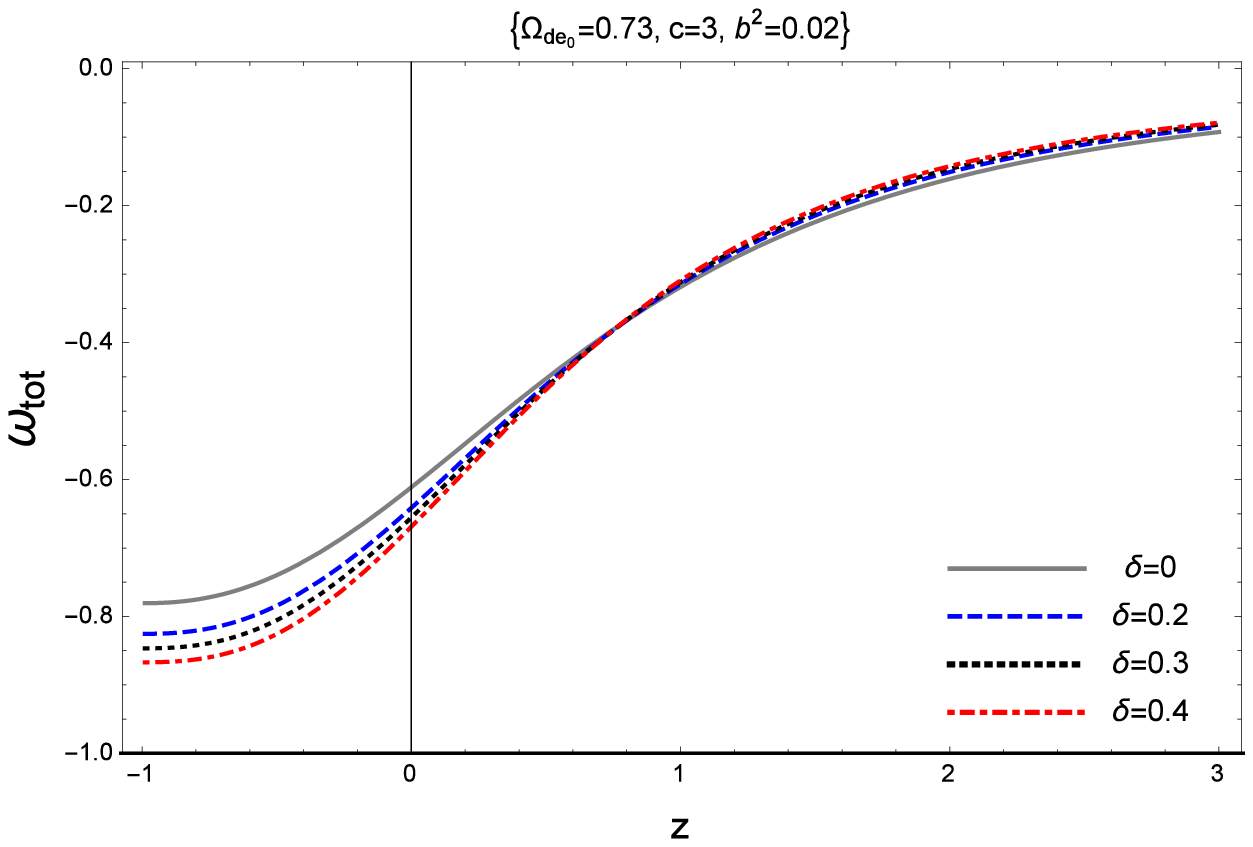}\text{(c)}
        \caption{The evolution of the $\Omega_D(z)$, $q(z)$ and $w_{tot}(z)$
            parameters for the original ADE in Barrow cosmology for different values of $\delta $ parameter.} \label{fig1}
    \end{center}
\end{figure}

\begin{figure}[htp]
    \begin{center}
        \includegraphics[width=8.7cm]{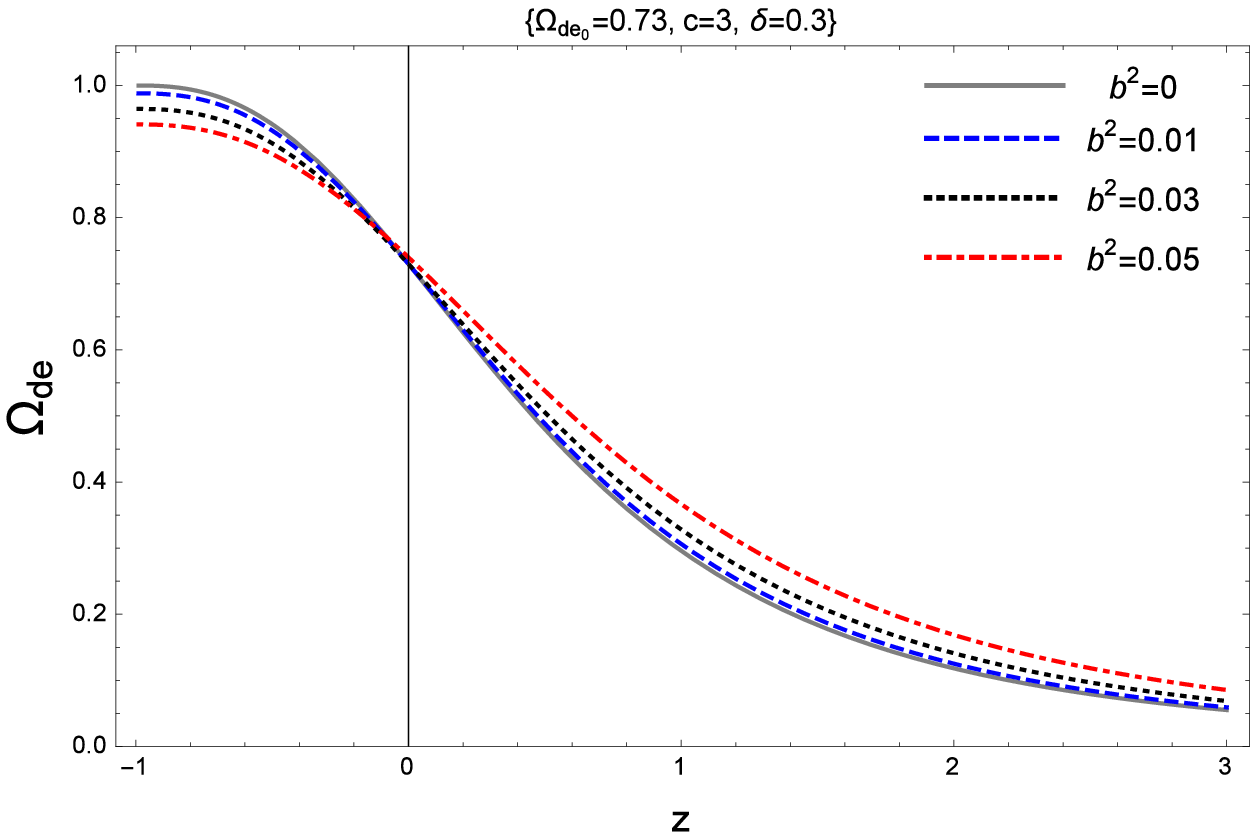}\text{(a)}
        \vspace{1mm}
        \includegraphics[width=8.7cm]{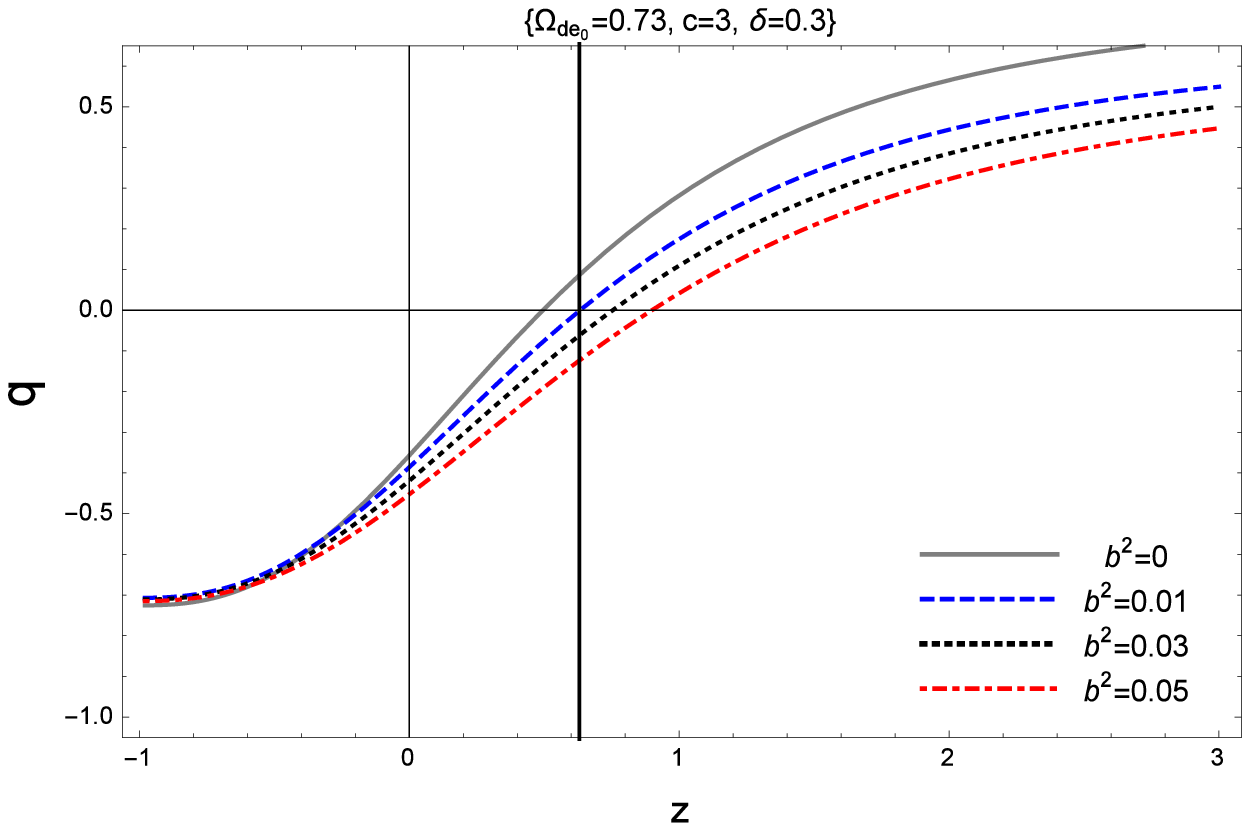}\text{(b)}
        \vspace{1mm}
        \includegraphics[width=8.7cm]{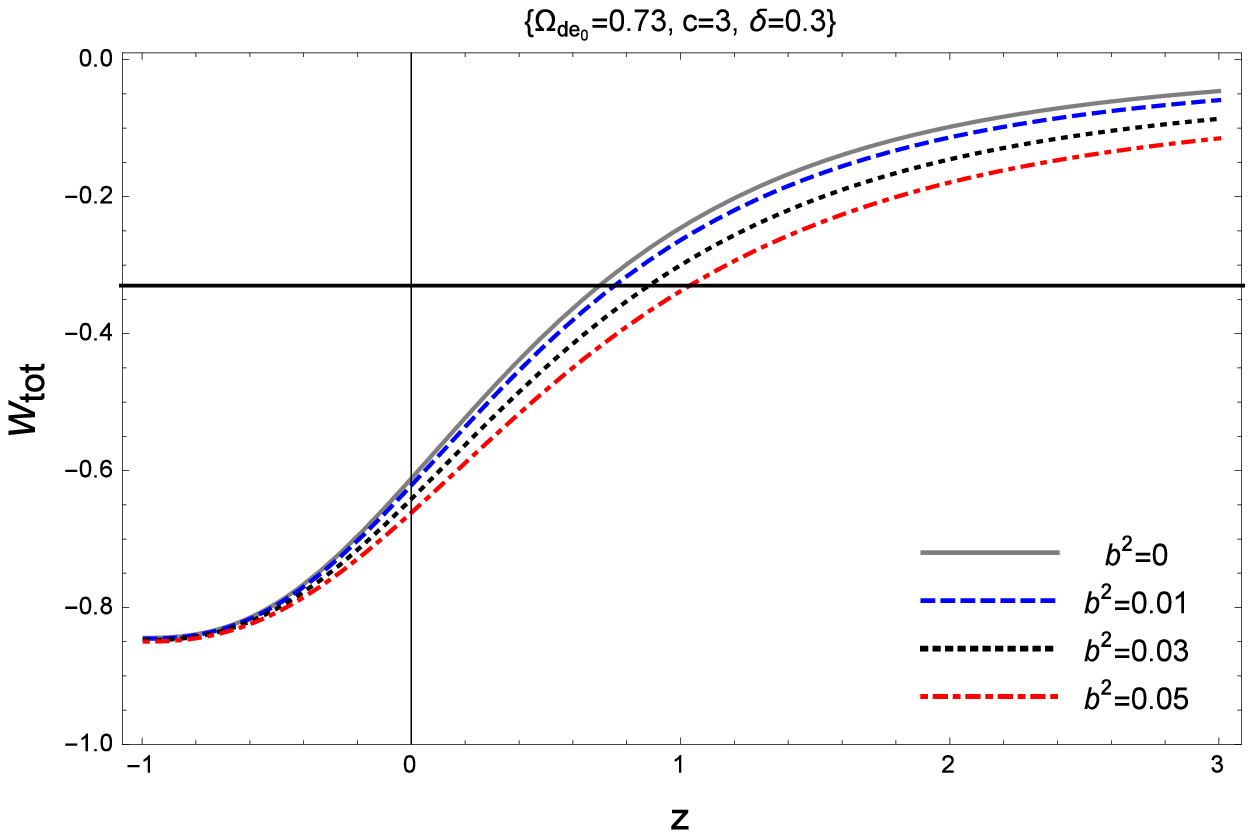}\text{(c)}
        \caption{The evolution of the $\Omega_D(z)$, $q(z)$ and $w_{tot}(z)$
            parameters for the original ADE in Barrow cosmology for different values of $b^2$ parameter.} \label{fig2}
    \end{center}
\end{figure}
\section{THE ORIGINAL ADE IN Barrow cosmology\label{ORI}}
If we consider a time dependent EoS parameter for DE, we need to
replace $\rho_{\Lambda}$ with $\rho_{de}$. In this section, we
consider ADE as a time varying candidate of DE. The energy density
of the original ADE is given by \cite{Cai1}
\begin{equation}\label{rho1} \rho_{de}=
\frac{3c^2 m_{p}^2}{T^2},
\end{equation}
where $c$ is a constant and $T$ is the age of the universe,
\begin{equation}
T=\int{dt}=\int_0^a{\frac{da}{Ha}}.
\end{equation}
Consider a flat FRW universe filled with pressureless matter and
ADE. The first Friedmann equation (\ref{Fried1B}) can be written
\begin{equation} \label{fried}
H^{2-\delta}=\frac{1}{3 M^{2}_{\rm eff}}(\rho_m+\rho_{de}),
\end{equation}
where we have defined $M^{2}_{\rm eff}= (8 \pi G_{\rm eff})^{-1}$.
When entropy is changed, the energy density of ADE get modified as
well. The modified energy density of ADE, inspired by Barrow
entropy, is given by \cite{Emm1} for
\begin{equation}\label{rhoB}
\rho_{de}=\frac{3c^2 M^{2}_{\rm eff}}{T^{2-\delta}}.
\end{equation}
For generality, we consider the interaction between DM and ADE. In
this case, they satisfy the semi-conservation equations,
\begin{eqnarray}
&&\dot{\rho}_m+3H\rho_m=Q, \label{consm}
\\&& \dot{\rho}_D+3H\rho_D(1+w_{de})=-Q,\label{consq}
\end{eqnarray}
where $w_{de}$ is the EoS parameter of ADE and  $Q =3b^2
H(\rho_{de}+\rho_m)$ is the interaction term, with $b^2$ being a
coupling constant.

Using Eq. (\ref{rhoB}) and $\rho_c=3 M^{2}_{\rm eff}H^{2-\delta}$,
the density parameter $\Omega_{de}$ can be written
\begin{eqnarray}\label{Omega1}
\Omega_{de}=\frac{\rho_{de}}{\rho_c}=\frac{c^2}{(HT)^{2-\delta}}.
\end{eqnarray}
Taking the time derivative of (\ref{rhoB}), we find
\begin{eqnarray}\label{rhodot}
\dot{\rho}_{de}=\frac{\delta-2}{T}\rho_{de}
\end{eqnarray}
Substituting (\ref{rhodot}) into (\ref{consq}) after using
relation (\ref{Omega1}), we find the EoS parameter of ADE in
Barrow cosmology,
\begin{eqnarray}\label{wde}
w_{de}=-1+\frac{2-\delta}{3}\left(\frac{\Omega_{de}}{c^2}\right)^{1/(2-\delta)}-b^2
\Omega^{-1}_{de}.
\end{eqnarray}
Taking the time derivative of  both side of the Friedman equation
(\ref{fried}) and using Eqs. (\ref{consq})  and (\ref{Omega1}), it
is easy to find that
\begin{eqnarray}\label{Hdot}
\frac{\dot{H}}{H^2}=-\frac{3}{2-\delta}(1+\Omega_{de}w_{de}).
\end{eqnarray}
Using (\ref{wde}), the above relation can be written
\begin{eqnarray}\label{Hdot2}
\frac{\dot{H}}{H^2}=-\frac{3}{2-\delta}(1-\Omega_{de})+\frac{3b^2}{2-\delta}-\Omega_{de}\left(\frac{\Omega_{de}}{c^2}\right)^{1/(2-\delta)}.\nonumber\\
\end{eqnarray}
Also, if we take the time derivative of (\ref{Omega1}), after
using (\ref{Hdot}) and $\dot\Omega_{de}=H {\Omega^\prime_{de}}$,
we find
\begin{eqnarray}\label{Omegaq2}
{\Omega'_{de}}=\Omega_{de}\Bigg{\{}3(1+\Omega_{de}w_{de})-(2-\delta)\left(\frac{\Omega_{de}}{c^2}\right)^{1/(2-\delta)}\Bigg{\}},\nonumber\\
\end{eqnarray}
where the prime denotes derivative with respect to $x=\ln{a}$.
Substituting $w_{de}$ from Eq. (\ref{wde}) into the above
relation, yields
\begin{equation}
{\Omega^\prime_{de}}=\Omega_{de}(1-\Omega_{de})\Bigg{\{}
3-(2-\delta)\left(\frac{\Omega_{de}}{c^2}\right)^{1/(2-\delta)}\Bigg{\}}-3
b^2 \Omega_{de}.\label{omegaprime2}
\end{equation}
The total EoS parameter is given by
\begin{eqnarray}\label{wtot}
&&w_{\mathrm{tot}}=\frac{p_{de}}{\rho_m+\rho_{de}}=\Omega_{de} w_{de}\nonumber\\
&&=\Omega_{de}\left[-1+\frac{2-\delta}{3}\left(\frac{\Omega_{de}}{c^2}\right)^{1/(2-\delta)}\right]-b^2.
\end{eqnarray}
For completeness, we give the deceleration parameter
\begin{eqnarray}
q=-\frac{\ddot{a}}{aH^2}=-1-\frac{\dot{H}}{H^2},\label{q0}
\end{eqnarray}
which combined with the Hubble parameter and the dimensionless
density parameters form a set of useful parameters for the
description of the astrophysical observations. Substituting Eq.
(\ref{Hdot}) in Eq. (\ref{q0}) we get
\begin{eqnarray}\label{q}
q&=&-1+\frac{3}{2-\delta}(1-\Omega_{de})+\Omega_{de}\left(\frac{\Omega_{de}}{c^2}\right)^{1/(2-\delta)}\nonumber\\
&&-\frac{3b^2}{2-\delta}.
\end{eqnarray}
We have plotted the evolutions of $\Omega_{de}$, $q$ and
$w_{\mathrm{tot}}$ versus redshift $z$ for interacting original
ADE in Barrow cosmology for different values of $\delta$ and $
b^2$ in Figs. \ref{fig1} and \ref{fig2}, respectively. And in all
Figs., the $ \delta=0 $ case is also considered so that the model
can be compared with the standard model. In all figures, we have
taken $\Omega^{0}_{de}=0.73$ in agreement with observations and
$c=3 $.

In figs. \ref{fig1}(a) and \ref{fig2}(a), it can be easily seen
that at the early universe, $(z\rightarrow \infty)$, we get $
\Omega_{de}\rightarrow 0 $, while at the late time $ (z\rightarrow
-1) $, we have DE dominated era; namely, $ \Omega_{de} \rightarrow
1 $.

We have shown the behavior of the deceleration parameter $ q(z) $
vs $ z $ in Figs. \ref{fig1}(b) and \ref{fig2}(b). From these
figures we find out that, in both cases, our Universe has a
transition from deceleration to the acceleration phase around $
0.6 < z <0.9 $ which is compatible with observations
\cite{Daly,Komatsu,Salvatelli}. Moreover, we observe that by
decreasing $ \delta $ (increasing $ b^2 $), the transition happens
earlier (higher redshift), while the value of the deceleration
parameter at the present time, $q_0$, decreases with increasing
$b^2$ (decreasing $\delta$). For $ \delta=0.3 $, the transition
occurs around $ z_{\rm trn}=0.63 $ which is the best fit  with
observational data. In general, as one can see the behavior of the
deceleration parameter for $ 0<\delta<0.5 $ is more consistent
with the cosmological observation than the $ \delta=0 $ case
(standard ADE).

In Figs.\ref{fig1}(c) and \ref{fig2}(c), we have plotted the
behavior the total EoS parameter, $ w_\mathrm{tot}(z)$, for the
different values of $\delta$ and $ b^2 $, respectively. It is
obvious that the total EoS parameter, according to the
cosmological observations, can explain the evolution of the
universe, as at the early times $ (z\rightarrow\infty) $ we have $
w_\mathrm{tot}\rightarrow 0 $, it means that the pressureless DM
is dominant, and then it gets the quintessence regime $( -1<
w_\mathrm{tot}<-1/3) $ from the present period to the late time.

\begin{figure}[htp]
\begin{center}
        \includegraphics[width=8cm]{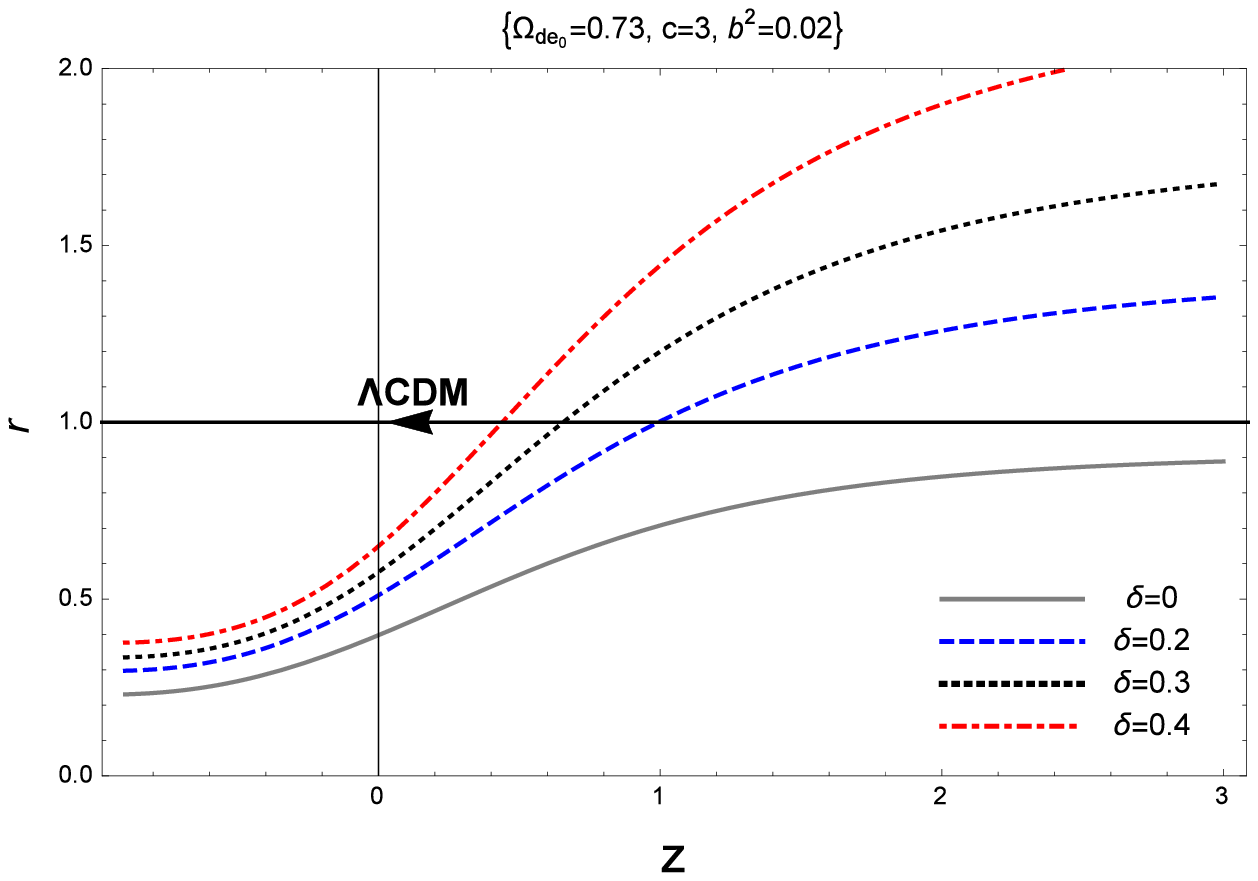}
        \vspace{1mm}
        \includegraphics[width=8cm]{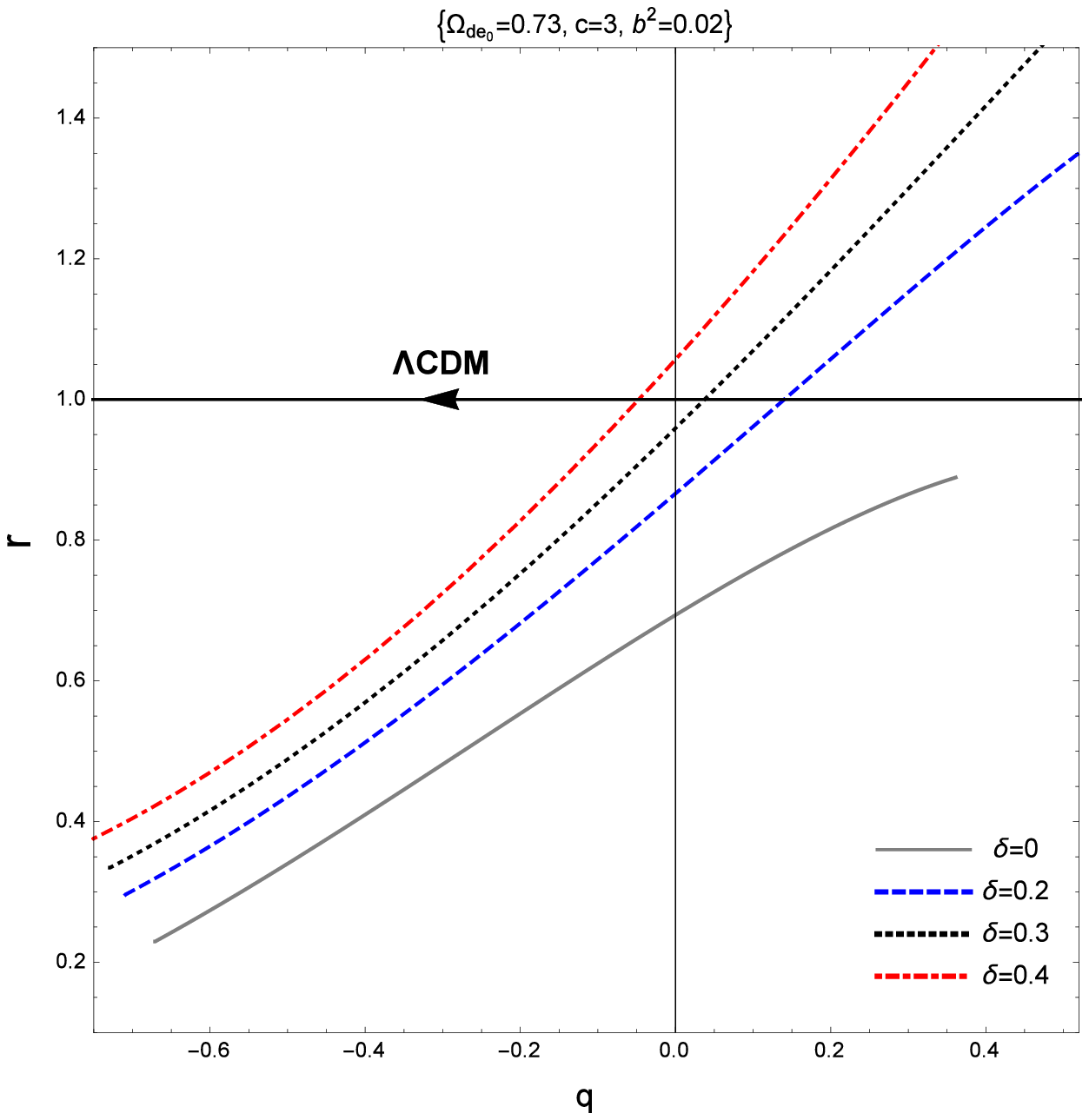}
        \vspace{1mm}
        \caption{The evolution of the $ r(z)$ and $r(q)$
         for original ADE in Barrow cosmology for different values of $ \delta $ parameter.} \label{fig3}
\end{center}
\end{figure}
\begin{figure}[htp]
\begin{center}
    \includegraphics[width=8cm]{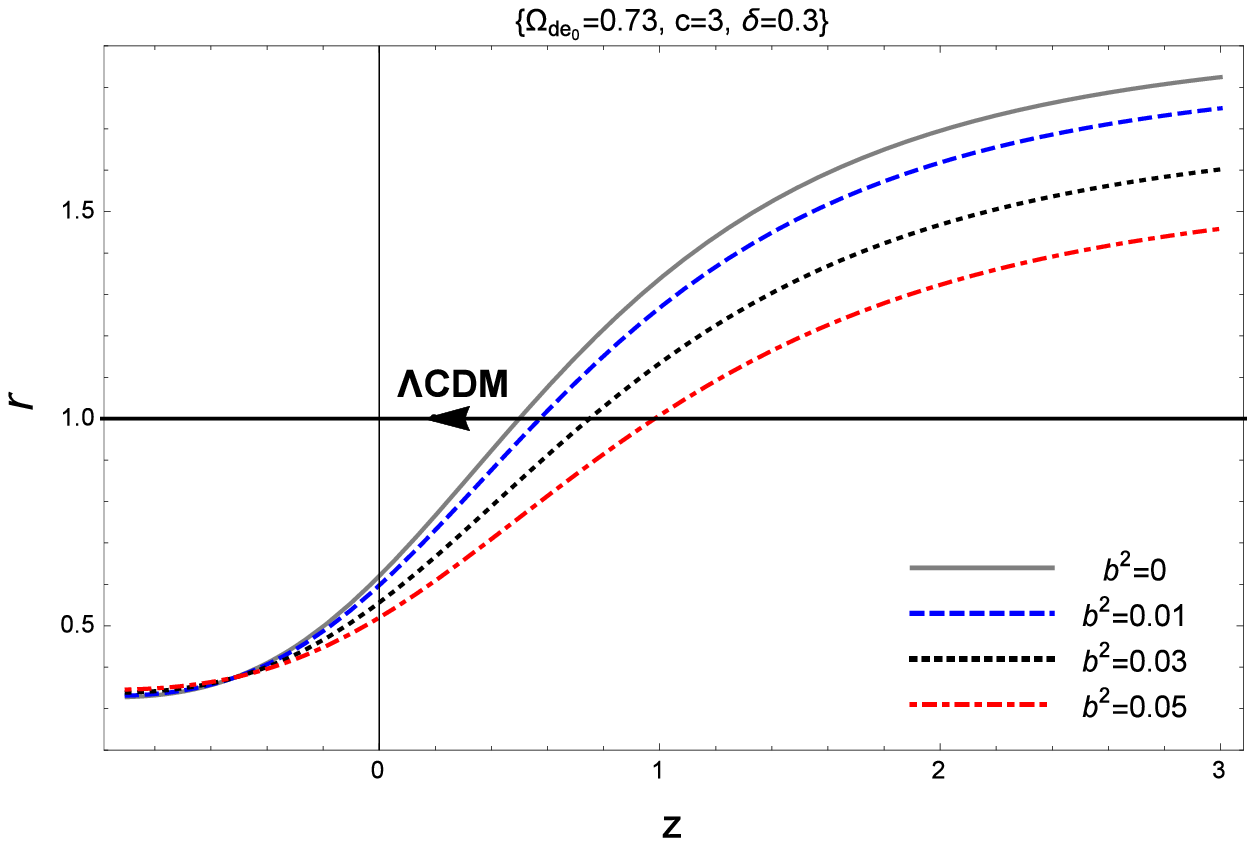}
    \vspace{1mm}
    \includegraphics[width=8cm]{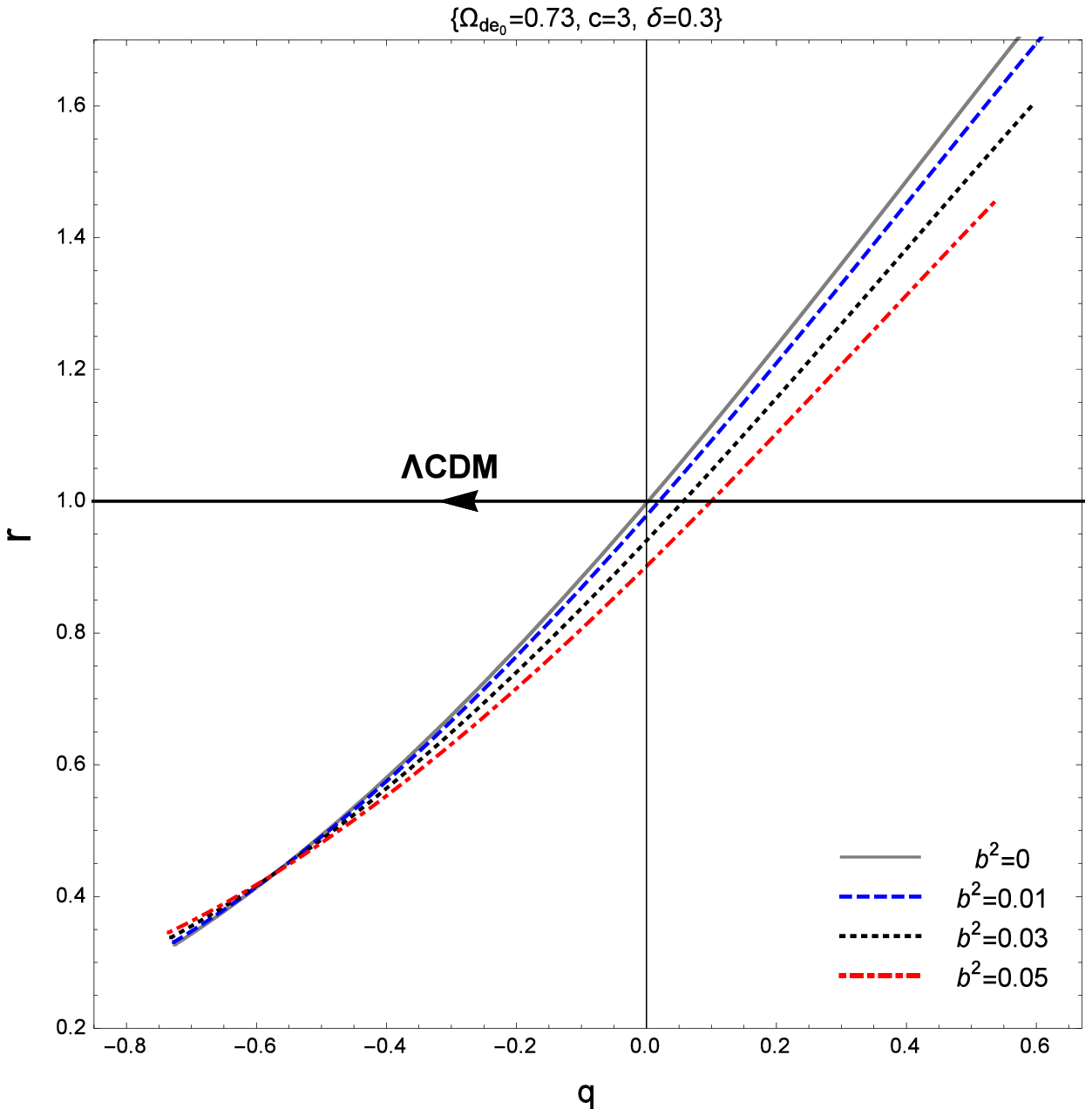}
    \vspace{1mm}
    \caption{The evolution of the $ r(z)$ and  $r(q)$
    for original ADE in Barrow cosmology for different values of $ b^2 $ parameter.} \label{fig4}
    \end{center}
\end{figure}
Furthermore, in order to find more sensitive discriminator of the
expansion rate, we consider the statefinder parameters which
contain the third derivative of the scale factor. The statefinder
pair $\{r,s\}$ is defined as \cite{GhaffariRevD}
\begin{eqnarray}\label{r}
&&r=\frac{\dddot{a}}{aH^3}=1+3\frac{\dot{H}}{H^2}+\frac{\ddot{H}}{H^3}\\
&&s=\frac{r-1}{3(q-1/2)}.
\end{eqnarray}
Since the statefinder parameters include the higher derivative of
the scale factor, it can be a fine tool to distinguish different
DE models. It is worth noting that the statefinder parameter $
\{r,s\}=\{1,0\} $ corresponds to the $ \Lambda CDM$ model and it
can be used to distinguish the difference between a given DE model
from the $ \Lambda CDM$  model in the $ (r-s) $ plane.

Taking the time derivative of both sides of Eq. (\ref{Hdot}) and
using Eq. (\ref{wde}), we obtain the statefinder pair $ \{r,s\} $
for the original ADE in Barrow cosmology as
\begin{eqnarray}\label{r2}
&&r=1-\frac{9}{2-\delta}\left(1+\Omega_{de}w_{de}\right)\left(1-\frac{2}{2-\delta}(1+\Omega_{de}w_{de})\right)\nonumber\\&&-\frac{3\Omega^\prime_{de}}{2-\delta}
\Biggl\{w_{de}+\frac{1}{3}\left(\frac{\Omega_{de}}{c^2}\right)^{1/(2-\delta)}+\frac{b^2}{\Omega_{de}}\Biggr\},
\end{eqnarray}
and
\begin{eqnarray}\label{s2}
&&s=\Biggl\{\frac{-9}{2-\delta}\left(1+\Omega_{de}w_{de}\right)\left(1-\frac{2}{2-\delta}(1+\Omega_{de}w_{de})\right)\nonumber\\&&-\frac{3\Omega^\prime_{de}}{2-\delta}
\Biggl[w_{de}+\frac{1}{3}\left(\frac{\Omega_{de}}{c^2}\right)^{1/(2-\delta)}+\frac{b^2}{\Omega_{de}}\Biggr]\Biggr\}\times\nonumber\\&&
\Biggl\{-\frac{3}{2}+\frac{3}{2-\delta}(1-\Omega_{de}-b^2)+\Omega_{de}\left(\frac{\Omega_{de}}{c^2}\right)^{1/(2-\delta)}\Biggr\}^{-1}.\nonumber\\
\end{eqnarray}
The evolutionary behavior of the statefinder parameter $ r(z)$ and
$ r(q) $ are plotted in Figs.~\ref{fig3} and \ref{fig4} for the
original ADE in Barrow cosmology for different values of the $
\delta  $ and  $ b^2 $ parameters, respectively. It should be
noted that in all figures, the black solid horizontal line denotes
the evolution of the $ \Lambda CDM $ mode.

From Figs. \ref{fig3} and \ref{fig4} we observe that the behavior
of the parameters $ r(z) $ and $ r(q) $ distinct from $ \Lambda
CDM $ model during the evolution of the universe.
\section{THE NEW MODEL OF ADE IN Barrow cosmology\label{NEW}}
\begin{figure}[htp]
    \begin{center}
        \includegraphics[width=8.7cm]{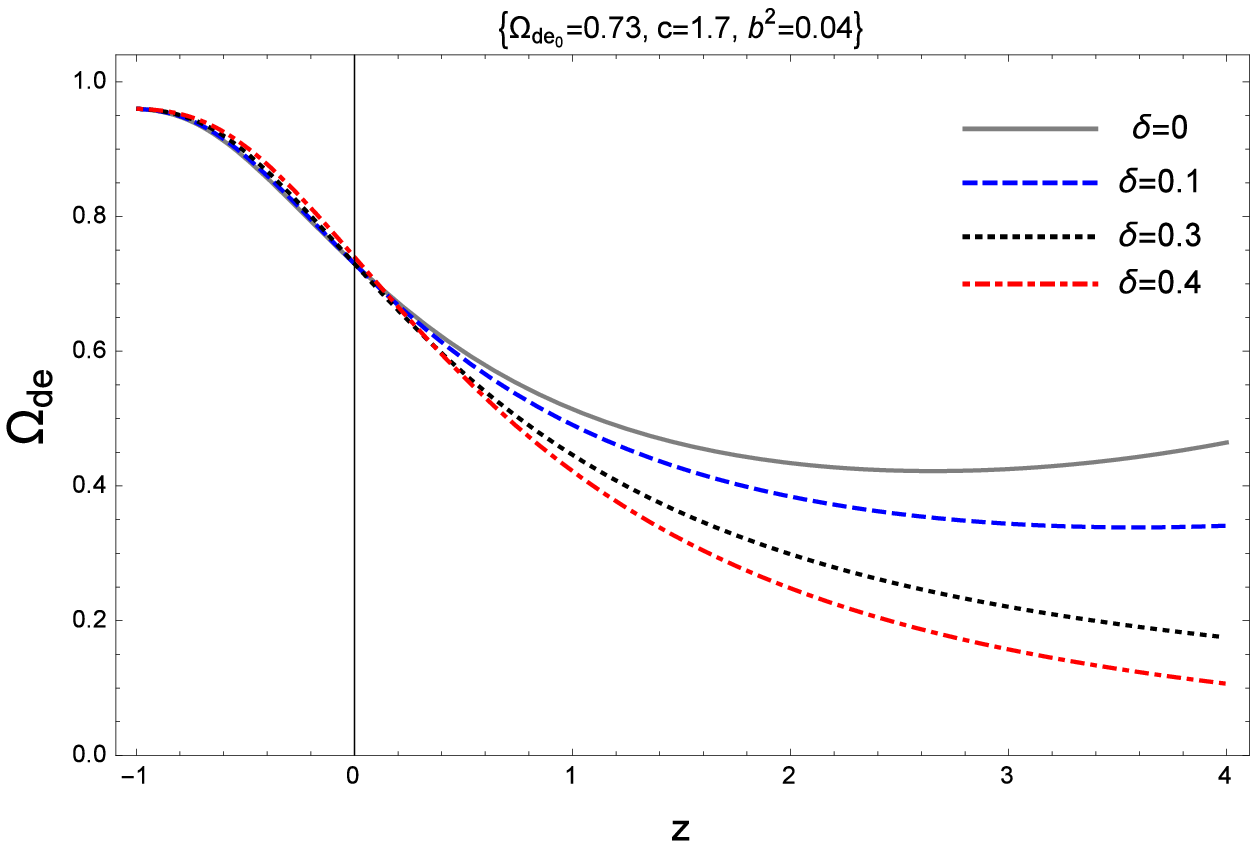}\text{(a)}
        \vspace{1mm}
        \includegraphics[width=8.7cm]{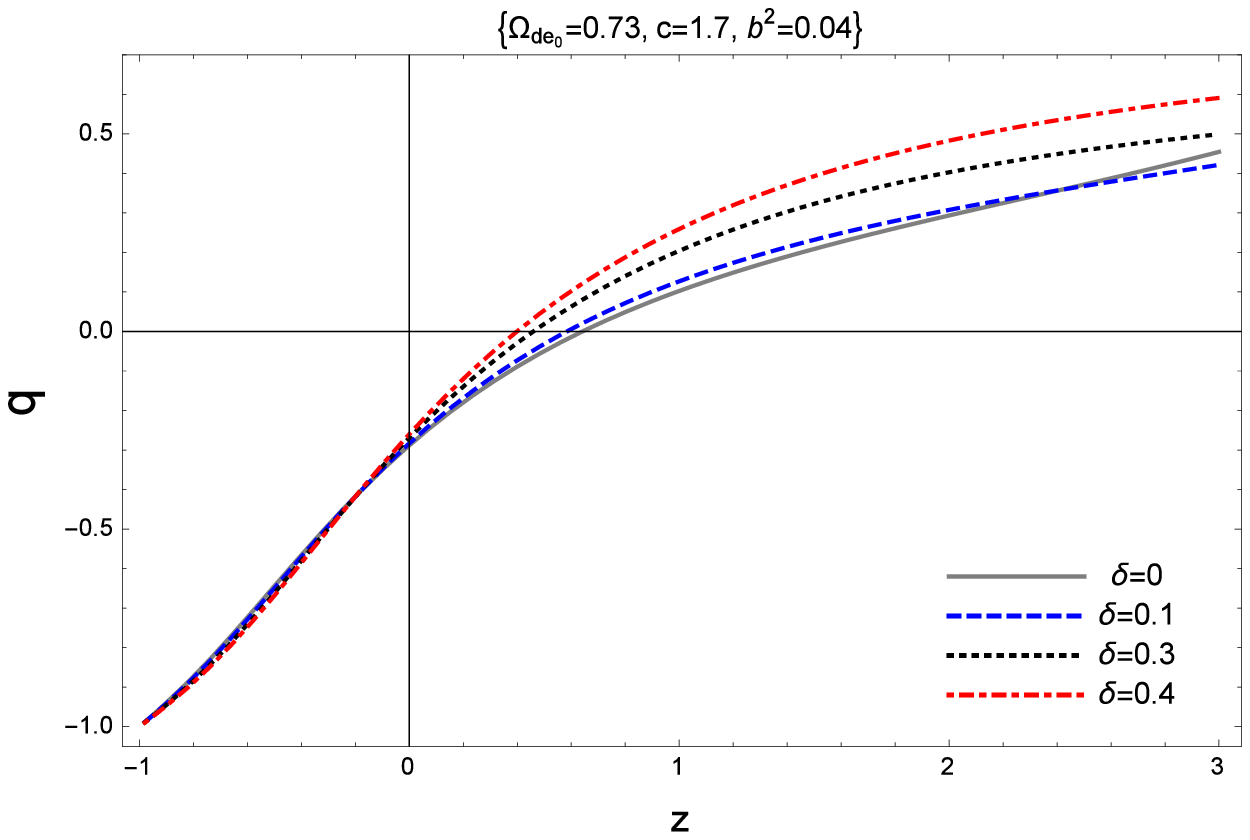}\text{(b)}
        \vspace{1mm}
        \includegraphics[width=8.7cm]{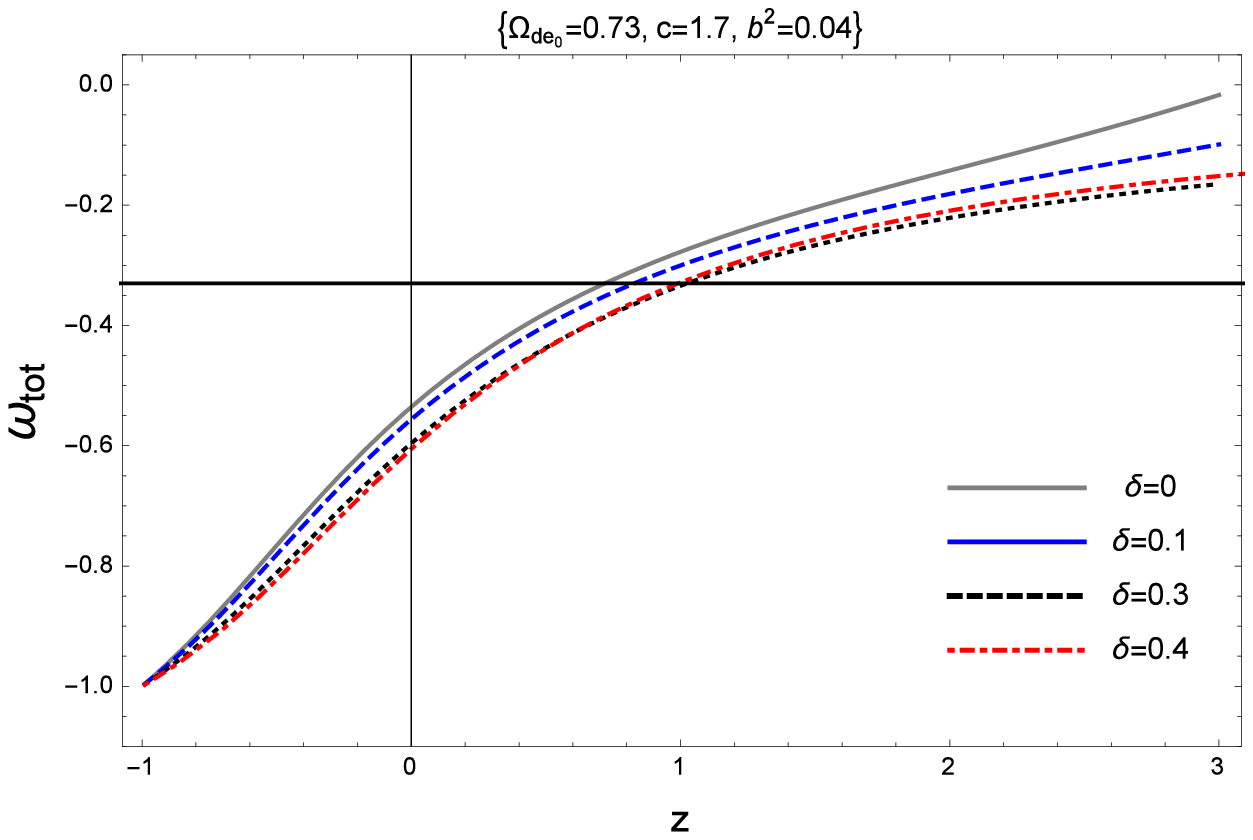}\text{(c)}
        \caption{The evolution of the $\Omega_D(z)$, $q(z)$ and $w_{tot}(z)$
           for the new ADE in Barrow cosmology for different values of $ \delta $ parameter.} \label{fig5}
    \end{center}
\end{figure}

\begin{figure}[htp]
    \begin{center}
        \includegraphics[width=8.7cm]{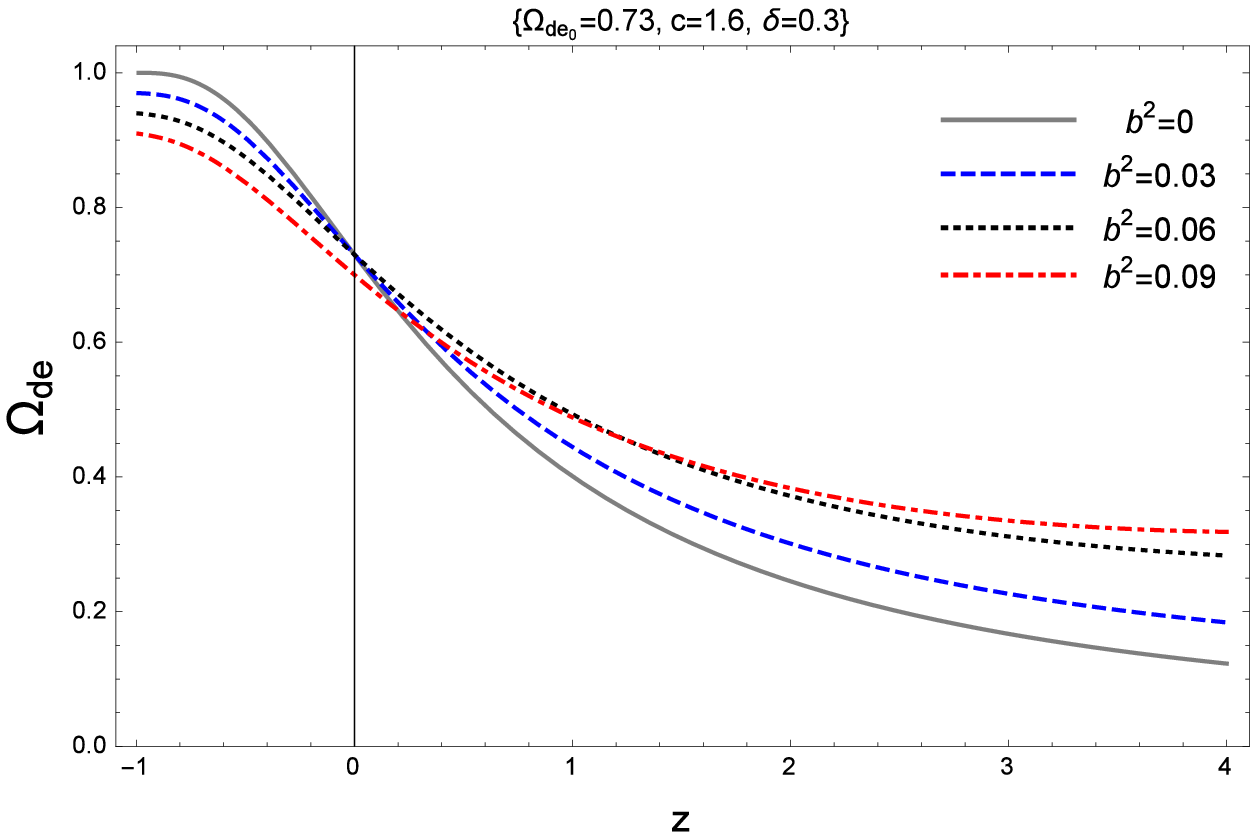}\text{(a)}
        \vspace{1mm}
        \includegraphics[width=8.7cm]{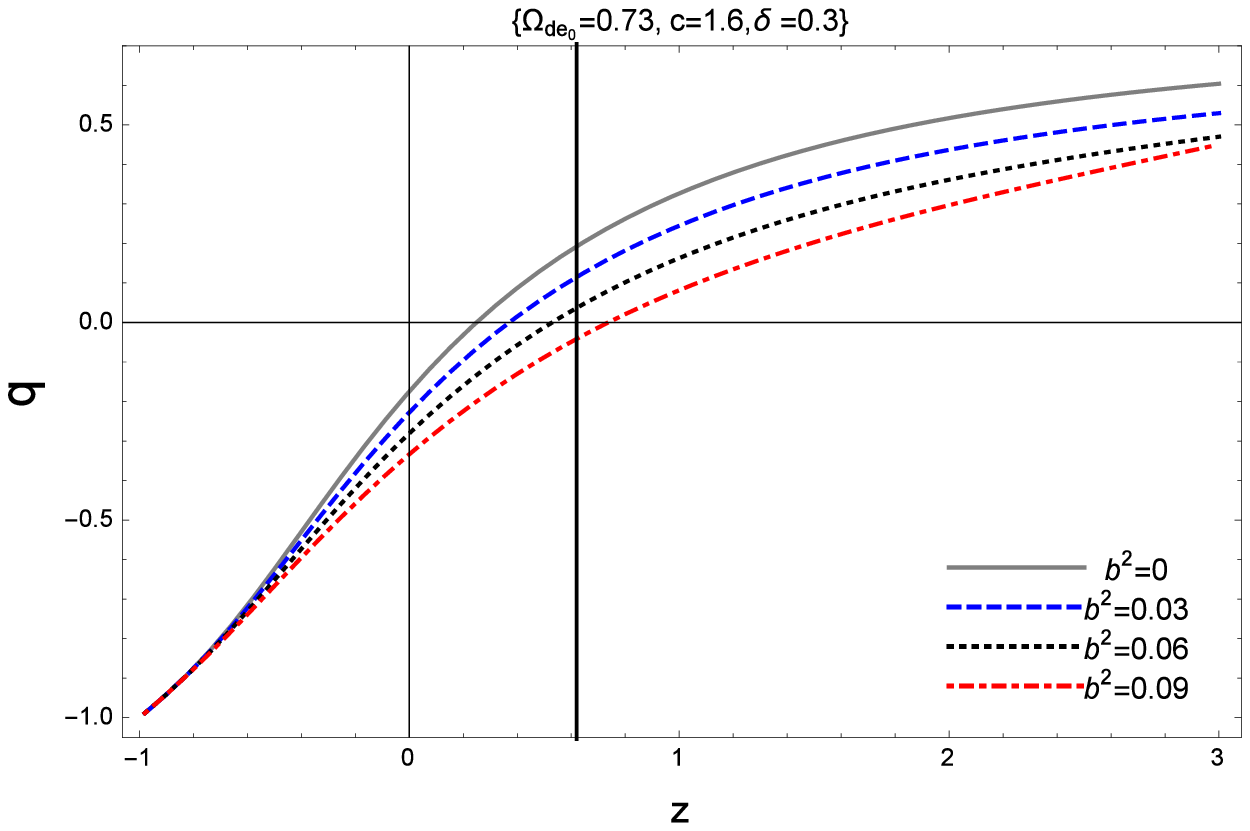}\text{(b)}
        \vspace{1mm}
        \includegraphics[width=8.7cm]{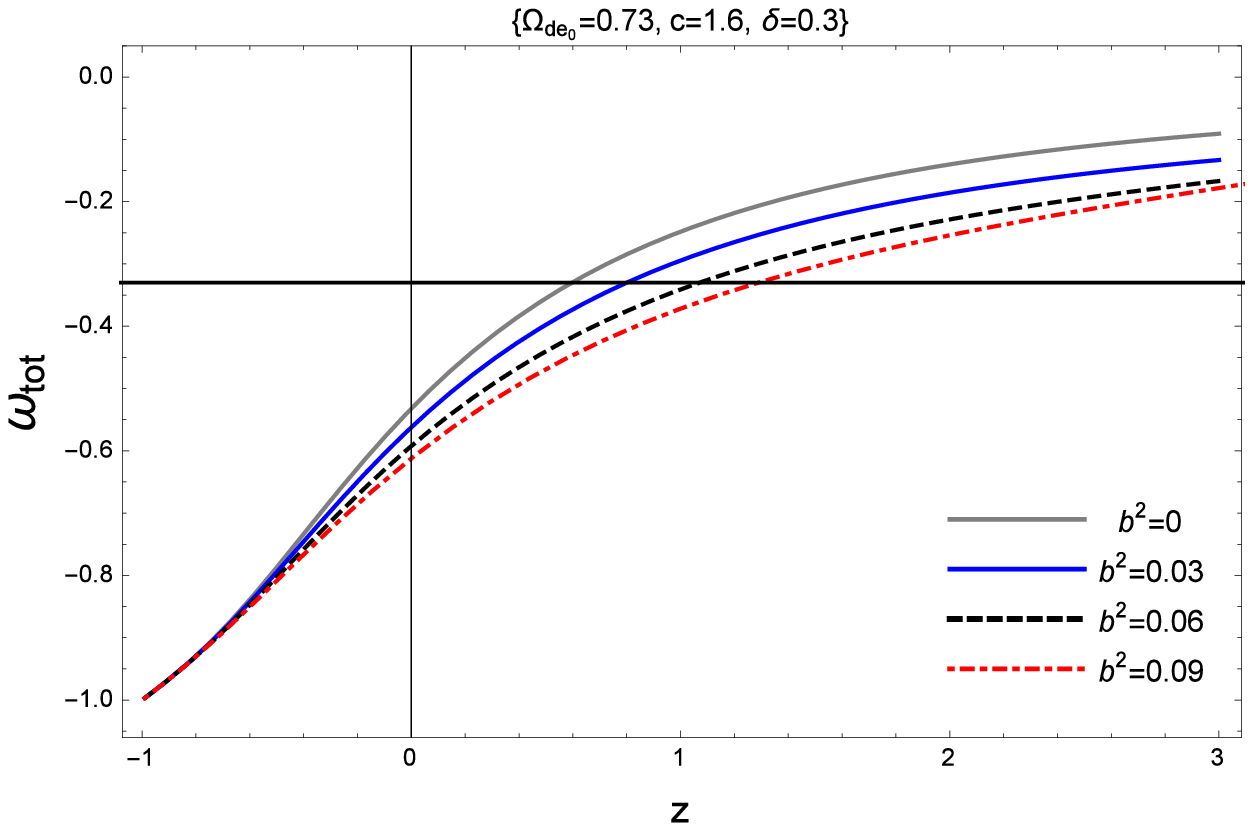}\text{(c)}
        \caption{The evolution of the $\Omega_D(z)$, $q(z)$ and $w_{tot}(z)$
           for the new ADE in Barrow cosmology for different values of $ b^2 $ parameter.} \label{fig6}
    \end{center}
\end{figure}

Another version of ADE has been also investigated in the
literatures where instead of the age of the universe, the
conformal time $\eta$ is chosen as IR cutoff. Taking this into
account, the energy density of the new ADE is given by
\begin{equation}\label{rho3}
\rho_{de}= \frac{3c^2 m_{p}^2}{\eta^2},
\end{equation}
where the conformal time is given by
\begin{equation}
\eta=\int{\frac{dt}{a}}=\int_0^a{\frac{da}{Ha^2}},
\end{equation}
and hence $\dot{\eta}=1/a$. Based on Barrow entropy and motivated
by (\ref{rho3}), we define the energy density of the new ADE in
Barrow cosmology as
\begin{equation}\label{rhoB2}
\rho_{de}=\frac{3c^2 M^{2}_{\rm eff}}{\eta^{2-\delta}},
\end{equation}
In order to study the cosmological implications of this model, we
first derive the EoS parameter. Taking the time derivative of
(\ref{rho3}) we find
\begin{eqnarray}\label{rhodot2}
\dot{\rho}_{de}=\frac{\delta-2}{\eta a}\rho_{de}.
\end{eqnarray}
The density parameter in this case reads
\begin{eqnarray}\label{Omega2}
\Omega_{de}=\frac{\rho_{de}}{\rho_c}=\frac{c^2}{(H\eta)^{2-\delta}}.
\end{eqnarray}
Substituting Eq. (\ref{rhodot2}) into the continuity equation
(\ref{consq}), after some calculations and using relation
(\ref{Omega2}) we reach
\begin{eqnarray}\label{wde2}
w_{de}=-1+\frac{2-\delta}{3a}\left(\frac{\Omega_{de}}{c^2}\right)^{1/(2-\delta)}-b^2
\Omega^{-1}_{de}.
\end{eqnarray}
In this case, it is a matter of calculations to show that the
evolutions of $\Omega_{de}$ and $q$ are governed by the following
equations
\begin{equation}
{\Omega^\prime_{de}}=\Omega_{de}(1-\Omega_{de})\Bigg{\{}
3-\frac{2-\delta}{a}\left(\frac{\Omega_{de}}{c^2}\right)^{1/(2-\delta)}\Bigg{\}}-3
b^2 \Omega_{de},\label{omegaprime3}
\end{equation}
\begin{eqnarray}\label{q}
&&q=-1+\frac{3}{2-\delta}(1-\Omega_{de})+\frac{\Omega_{de}}{a}\left(\frac{\Omega_{de}}{c^2}\right)^{1/(2-\delta)}-\frac{3b^2}{2-\delta}.\nonumber\\
\end{eqnarray}
In Figs. \ref{fig5} and \ref{fig6}, we have plotted the evolutions
of $\Omega_{de}$, $q$ and $w_{\mathrm{tot}}$ versus $z$ for
interacting new ADE in Barrow cosmology where we fix $
\Omega_{de}(z=0) =\Omega_{de_0}=0.73$ (in agreement with
observations), $ c=1.7 $.

The evolution of the dimensionless density parameter $ \Omega_{de}
(z)$ is shown in Figs. \ref{fig5}(a) and \ref{fig6}(a) for new ADE
model. These figures confirm that we have a DM dominated universe
$ (\Omega_{de}\rightarrow 0) $ at the early stages of the
universe, and the DE dominated universe $ (\Omega_{de}\rightarrow
1) $ at the late time which is consistent with the cosmological
observations.

The behavior of the deceleration parameter $ q(z) $ is depicted in
Figs. \ref{fig5}(b) and \ref{fig6}(b). As one can see, the
universe undergoes a phase transition from the deceleration to an
acceleration within the interval(around) $ 0.7<z_{tr}<1 $. We
observe that for larger values of $ \delta $ parameter (smaller
values of $b^2 $), the phase transition occurs at the lower
redshifts.  We also find out that the deceleration parameter for
new ADE model in Barrow cosmology tends to $ -1 $ at the late time
and this is consistent with a de-Sitter expansion phase.

The behavior of the total EoS parameter are shown in Figs.
\ref{fig5}(c) and \ref{fig6}(c), for the different values of
$\delta$ and $ b^2 $, respectively. From these figures we see that
at the early times, we have $ w_{\mathrm{tot}}\rightarrow 0 $
which means that the universe is in the pressureless DM dominated
era while at the late time, we have $ w_{\mathrm{tot}}\rightarrow
-1 $ indicates the ending of the universe is in a Big-Rip
singularity.

\begin{figure}[htp]
    \begin{center}
        \includegraphics[width=8cm]{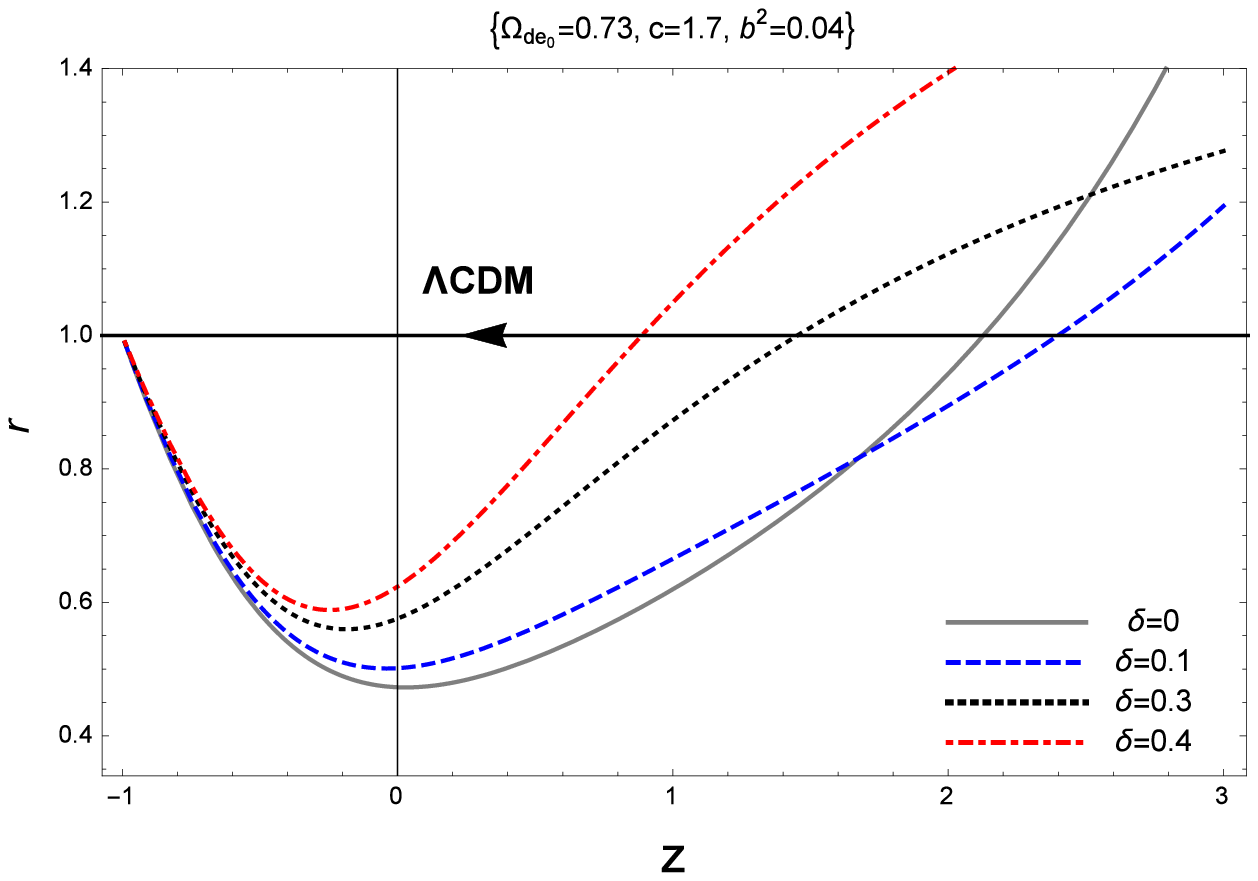}
        \vspace{1mm}
        \includegraphics[width=8cm]{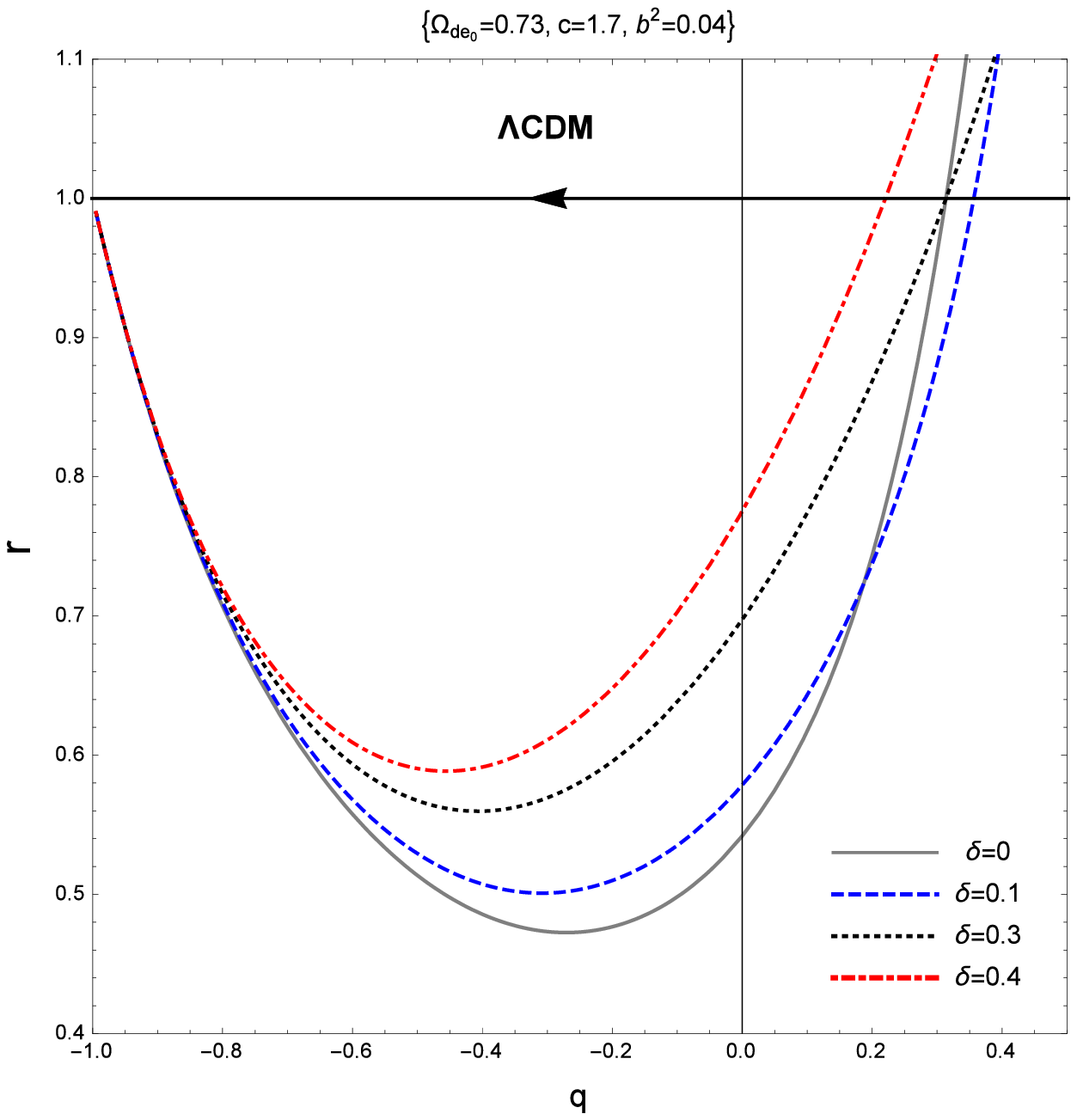}
        \vspace{1mm}
        \includegraphics[width=8cm]{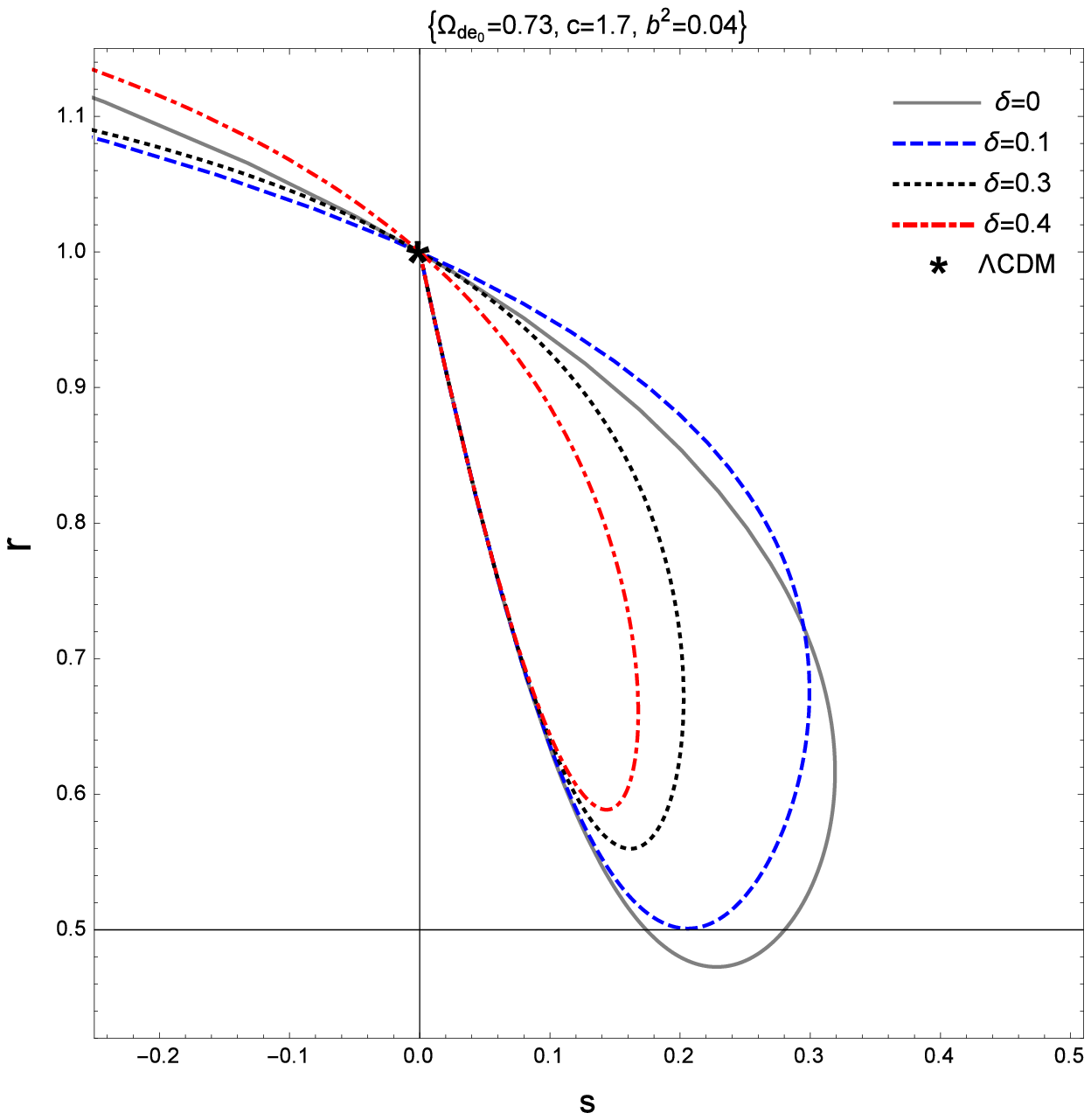}
        \caption{The evolution of the $ r(z) $ $r(q)$ and $r(s) $
         for the new ADE in barrow cosmology for different values of $\delta $ parameter.} \label{fig7}
    \end{center}
\end{figure}

\begin{figure}[htp]
\begin{center}
        \includegraphics[width=8cm]{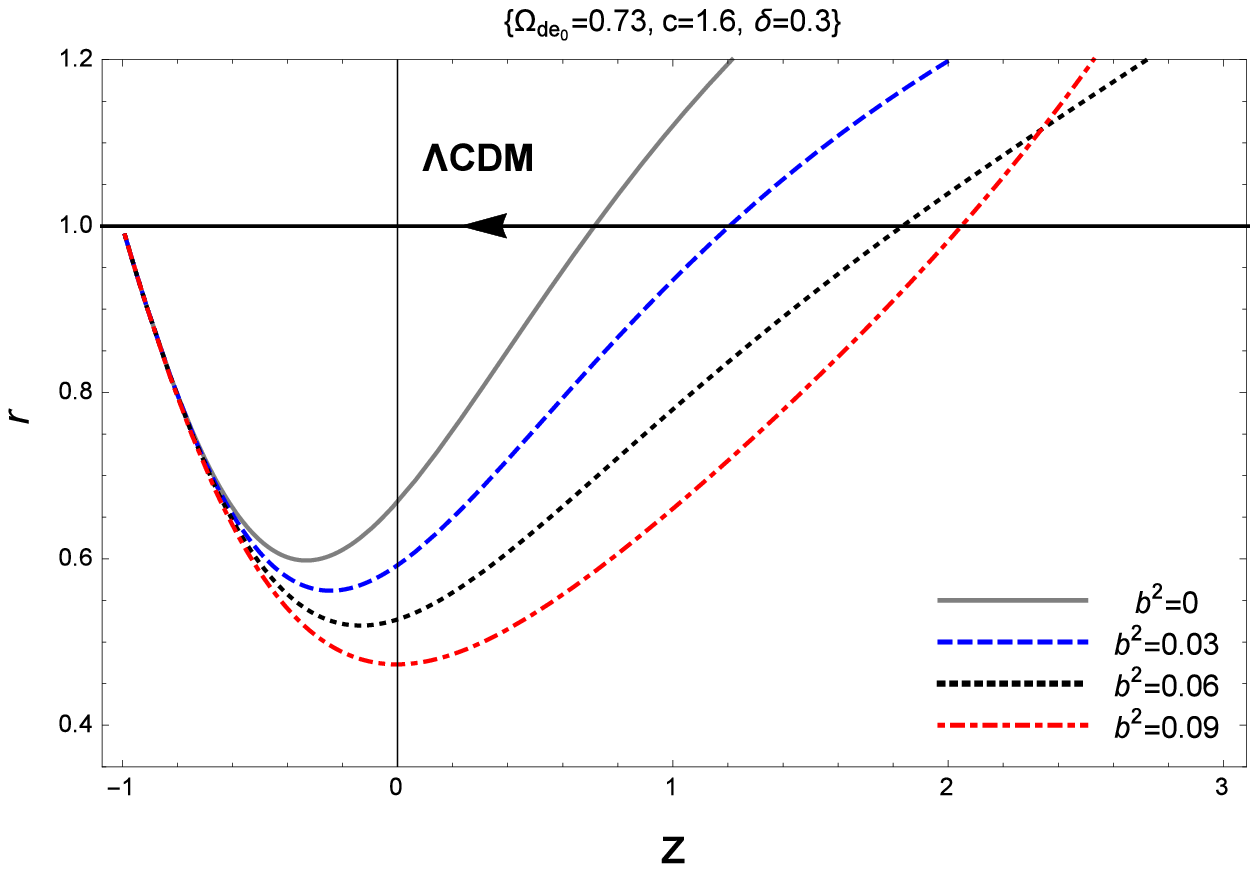}
        \vspace{1mm}
        \includegraphics[width=8cm]{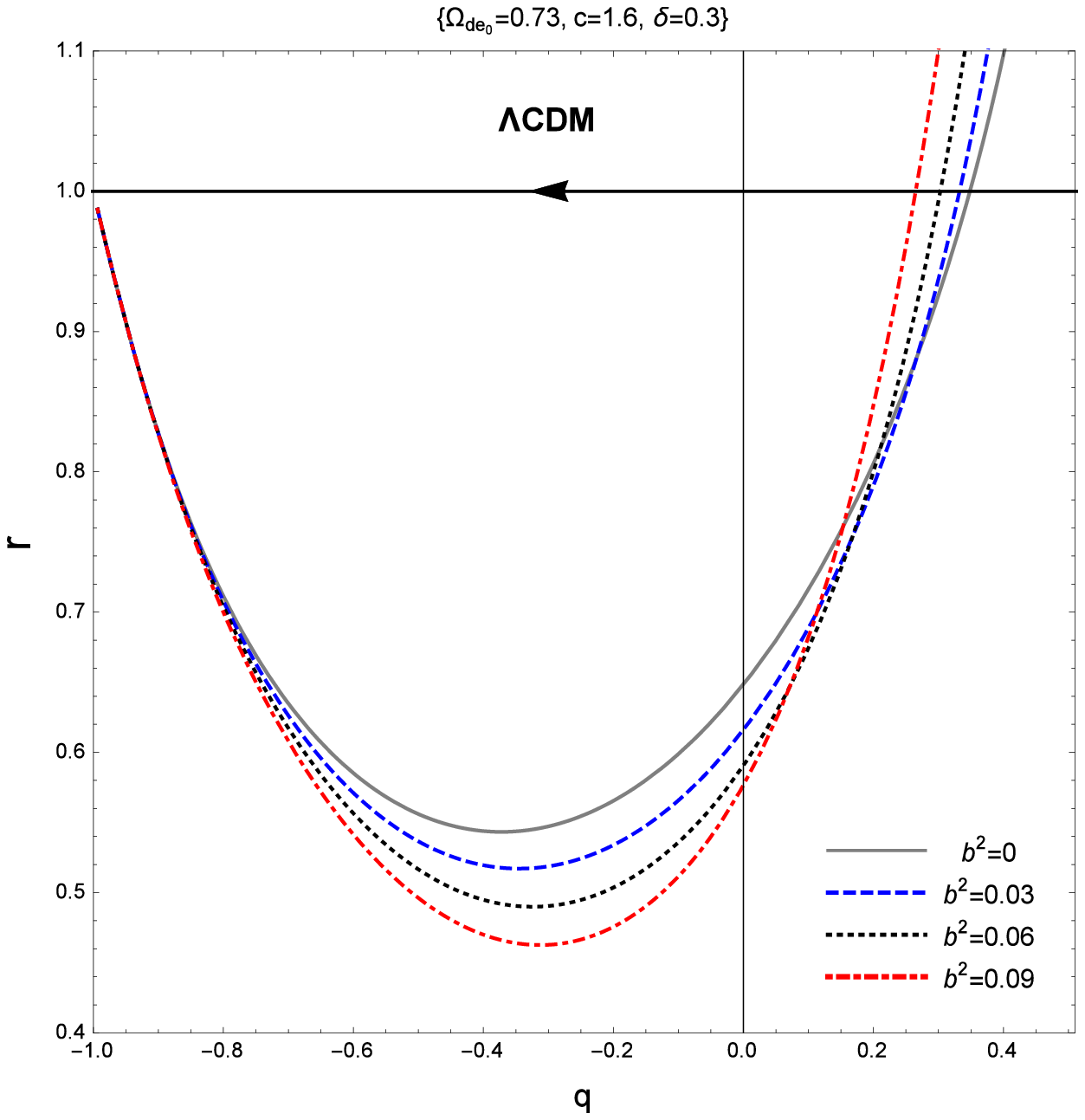}
        \vspace{1mm}
        \includegraphics[width=8cm]{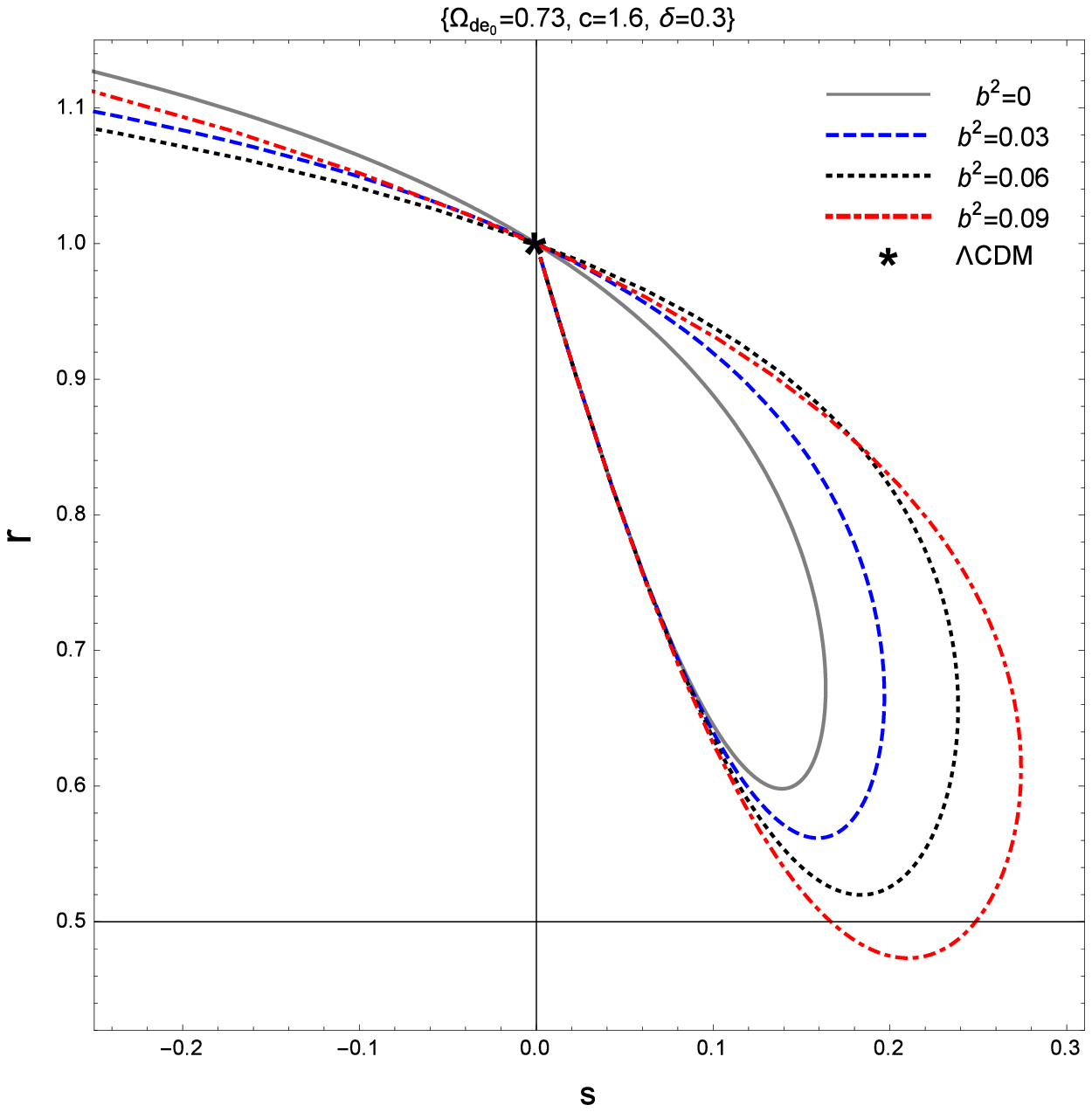}
        \caption{The evolution of the $ r(z) $ $r(q)$ and $r(s) $
       for the new ADE in barrow cosmology for different values of $ b^2 $ parameter.} \label{fig8}
    \end{center}
\end{figure}

By taking the time derivative of Eq. (\ref{Hdot}) and using Eq.
(\ref{wde2}), one can obtain the statefinder pair  $\{r,s\}$ for
the new ADE model in the Barrow cosmology. Since this expression
is too long, for the economic reason, we shall not present it
here, and only plot its evolution in Figs. \ref{fig7} and
\ref{fig8}.
We have also studied the behavior of $ r(z) $, $ r(s) $ and $ r(q)
$, for the new ADE in Barrow cosmology and for different values of
$ \delta$ and  $ b^2 $ parameters in Figs. \ref{fig7} and
\ref{fig8}, respectively. It should be noted that in $ r(z) $ and
$ (r-q) $ figures, the black solid horizontal line denotes the
evolution of the $ \Lambda CDM$ mode and the fixed point presented
by the star in $r-s $ panel is the $ \Lambda CDM$ model. From
these figures we find that the evolutionary behavior of the first
statefinder parameter $ r(z) $ and $(r-q) $, for new ADE in Barrow
cosmology for different values of the model parameters, is well
distinct from $ \Lambda CDM $ model in the region $ -0.5<z<3 $,
but it can not be distinguished from $ \Lambda CDM $ model in the
region $ -1<z<-0.5 $. In Figs. \ref{fig7} and \ref{fig8}, $( r-q
)$ plans show that both $ \Lambda CDM $ model and the new ADE
model start evolving from the same point $ (r,q)=(1,0.5) $ at the
early time, which indicate the DM dominated area, and they tend to
the $ (r,q)=(1,-1) $ in the future which correspond to the
de-Sitter expansion.

Finally, from the $r-s $ plans, one can observe that the
evolutionary trajectories of $(r-s) $ start from $ (r>1,s<0) $ in
the past, then it enters the quintessence region, i.e. $ (r<1,s>0)
$ at the present time and finally their evolution ends at the
$\Lambda CDM$ point $ (r,s)=(1.0) $ in the future for different
values of the model parameters.

\section{Conclusions and discussions\label{CONC}}
Inspired by the modified Barrow entropy, we have revisited the ADE
model in the background of Barrow cosmology. We have explored the
effects of the Barrow parameter $ \delta $ on the evolution of the
cosmological parameters for both original and new model of ADE. To
this end, we derived the differential equation for the evolution
of dimensionless ADE density parameter and used its numerical
solution in order to study the behavior of the corresponding
cosmological parameters. In all graphs we have set initial
condition $ \Omega_{de}(z=0) =\Omega_{de_0}=0.73$ in agreement
with observations.

From the behavior of deceleration parameter $ q $, we have found
that our universe, for both original ADE model an new ADE model in
the Barrow cosmology, has a phase transition from early
decelerated phase to the present accelerated phase for all choices
of the model parameter $ \delta $. For $ \delta=0.3 $ this
transition happens around  $ z_{\rm tr}\approx 0.63 $ which shows
a good compatibility with recent studies $0.5 < z_{\rm tr}<1 $. We
have also observed from evolution of the $ q(z) $ panel in Figs.
\ref{fig1} and \ref{fig5} that for smaller values of $ \delta $
parameter, transition from decelerated to accelerated phases
occurs at the earlier universe (higher redshifts). It is
noteworthy that for $ 0<\delta<0.5$ the behavior of the
deceleration parameter, for both original and new ADE models, is
more consistent with the cosmological observation than $ \delta=0
$ (standard ADE model).

We observed that the total EoS parameter $ w_{\rm tot} $ starts
from DE dominated universe( $ w_{\rm tot}= 0 $) at the early
universe($z\rightarrow \infty $) then enters to the quintessence
area ($ -1<w_{\rm tot}<-1/3 $) around the present time and finally
converges to the cosmological constant( $ w_{\rm tot}=-1 $ ) at
the late time independent of the value of $\delta$ parameter.

In the end, for the more clear discrimination among these DE
models, we used the statefinder pair, which include a higher-order
time derivative of the scale factor. By study the behavior of the
$r(z)$, $r-q $ and $ r-s $, we have found out that both original
ADE and new ADE models, are completely distinguishable from the $
\Lambda CDM $ model during the evolution of the universe. As we
have observed in Figs. \ref{fig7} and \ref{fig8}, for the new ADE
in Barrow cosmology, all trajectories of  $r-q $ and $r-s $, start
from the DM dominated at the early universe then enter to the DE
dominated universe (quintessence region) and finally end to the $
\Lambda CDM$ at the future time.


\end{document}